\documentstyle[12pt,aaspp4]{article}	

\def\Section#1{\S #1}

\newcommand{\etal}{\mbox{\it et~al.\ }}

\newcommand{\msun}{\mbox{$M_{\odot}$}}

\newcommand{\lsun}{\mbox{$L_{\odot}$}}
\newcommand{\kms}{\mbox{km s$^{-1}$}}

\def\deg      {{\ifmmode^\circ\else$^\circ$\fi} } 

\def\h2     {H$_2$}
\def\astar  {{Sgr A$^{*}$}}

\begin{document}

\title{HST P$\alpha$ and 1.9$\mu$m Imaging of Sgr A West}

\author{N. Z. Scoville}
\affil{Astronomy Department, California Institute of Technology, Pasadena, CA 91125}

\author{S. R. Stolovy }
\affil{SIRTF Science Center, California Institute of Technology, Pasadena, CA 91125}

\author{M. Rieke}
\affil{Steward Observatory, University of Arizona, Tucson, AZ 85721}

\author{M. Christopher}
\affil{Astronomy Department, California Institute of Technology, Pasadena, CA 91125}

\author{and F. Yusef-Zadeh}
\affil{Department of Physics and Astronomy, Northwestern University
         Evanston, IL 60208}

\begin{abstract}
We present HST/NICMOS images at 0.2\arcsec\ resolution (Nyquist sampled in 
the central 19\arcsec\ ) of the HI P$\alpha$ emission line in a 70\arcsec\ ($\alpha$) $\times$ 90\arcsec\ ($\delta$) region of the Galactic center centered on the non-thermal radio source Sgr A$^*$. The majority of the  emission arises from ionized gas in the mini-spiral in the central parsec.
P$\alpha$ emission is also seen from 26 stellar sources, presumably early-type  stars with mass-loss winds. The new data reveal 
significant small-scale structure ($\leq 1$\arcsec\ $\sim 1.2\times10^{17}$ cm) in the ionized gas of the mini-spiral; low 
surface brightness emission features are also seen for the first time. The ratio of observed P$\alpha$ emission to  
6-cm radio continuum emission is used to derive an extinction map for the ionized gas
over the central parsec. A$_V$ varies from 20 to 50 mag with a median value 
of A$_V = $ 31.1 mag, in excellent agreement with earlier estimates derived for the stellar sources. Extinctions, independently derived using H92$\alpha$ recombination line data, were in excellent agreement with those derived from the 6-cm continuum.  A broad minimum in the extinction extends $\sim$ 30\arcsec\ NE--SW over the area of the IRS 16 cluster, including the area of peak 
flux in P$\alpha$ and radio continuum. Large increases in extinction are seen along the periphery of the ionized gas, particularly in the 
direction of the western arm, suggesting that the ionized gas is {\it partially} extincted by dust in the molecular clouds at the outside of the ionized region;
however the HII cannot be entirely behind the molecular clouds since the extinctions 
would then be much greater. The minimum Lyman continuum emission rate within 40\arcsec\ radius of \astar ~is 3.9$\times 10^{50}$ sec$^{-1}$ or L$_{LyC} \simeq 2.7\times 10^6$ \lsun, half of this within 20\arcsec\ radius. The small-scale, filamentary structures in the ionized gas have a free thermal 
expansion time of only $\sim 3000$ yrs; either magnetic fields or mass-loss winds from 
the hot emission line stars may contain the ionized
filaments. (Eight new emission-line stars were also detected in P$\alpha$.) 

For both the
ionized gas and stellar continuum, the centroids of the emission (corrected for extinction) remain within $\sim$ 
$\pm$ 1\arcsec\ from a radius of 2\arcsec\ out to 40\arcsec\. This provides 
further evidence that \astar (the putative central black hole) is
indeed at or extremely close ($\leq 0.04$ pc or $1.2\times10^{17}$ cm) to the center of the Galactic nucleus stellar distribution and presumably the 
dynamical center of the Galaxy. The 1.9$\mu$m surface brightness increases
inwards to 0.9\arcsec\ radius and exhibits a decrease or leveling off 
closer to \astar, possibly indicating the core radius of the 
central stellar distribution or depletion of the late-type stars
by stellar collisions near the central black hole.

\end{abstract}
\keywords{Galaxy: center --- dust, extinction --- infrared: ISM  --- HII regions}


\section{Introduction}

Within the central few parsecs of our Galaxy, radio, infrared and 
x-ray observations have revealed a massive black hole lying within a region of ionized 'spiral' arms, encompassed by a ring of dense molecular clouds.   
The black hole is identified with the non-thermal radio source Sgr A$^*$,
and it is estimated to be 2.6$\times 10^6$ \msun\ from the motions of nearby stars and the neighboring 
ionized gas (\cite{eck96}, \cite{gen00}, \cite{ghe00}). At a projected radius of 
approximately 2 pc from \astar, lies the circumnuclear disk/ring (CND) of dense molecular gas and 
dust clouds (Genzel \etal 1985, G\"usten \etal 1987), and within 
this ring, ionized gas streams extend down to a few arcsec from Sgr A$^*$ (Lo \& Claussen 1983; Lacy \etal 1991). The molecular cloud ring 
can be modelled as a torus (0.5 pc thick) inclined at 70\deg to the line of sight (\cite{jac93}). Although clearly much 
less energetic and massive than the luminous AGN seen in many galactic nuclei,
this region probably provides our best opportunity to observe in detail the 
processes associated with buildup of massive black holes and their 
accretion processes.

Previous studies of the ionized gas in the inner few parsecs of the Galaxy, with an 
emphasis on the radial velocity kinematics have included observations of [NeII] 
(Lacy, Achtermann, \& Serabyn 1991),  Br$\gamma$ ( Burton \& Allen 1992 and 
Herbst, Beckwith, \& Shure 1993), and H92$\alpha$ (Roberts \& Goss 1993).
Roberts \& Goss published a ratio map of their H92 $\alpha$ radio observations to Br$\gamma$ from Burton \& Allen (1992) which gave a qualitative representation of the extinction in the inner few parsecs, however, they did not calculate the extinction in A$_V$.

We have recently undertaken two investigations aimed at providing the
highest resolution imaging of both the ionized and molecular gas in 
the circumnuclear disk. Most of the existing observations of the ionized and dense molecular gas are at $\geq$ 1 and $\geq$ 8 \arcsec\ resolution, respectively,
and thus do not clearly resolve the internal structure of the ionized gas.  
The molecular gas clouds were imaged in HCN at 
3\arcsec\ resolution by Christopher \etal (2003) using the Owens Valley 
mm-Array; here we present HST-NICMOS imaging at 0.2\arcsec\ resolution of the 1.87$\mu$m P$\alpha$ (HI 4--3) recombination emission line. The P$\alpha$ line has 12 times greater
intrinsic flux than the 2.16$\mu$m Br$\gamma$ line, and thus it provides 
excellent sensitivity when observed with the NICMOS camera on HST (despite the extra magnitude of extinction at the Galactic center at 1.87$\mu$m compared with 2.16$\mu$m). Comparison of the P$\alpha$ emission with radio continuum emission from the same ionized gas is used to derive a map of the foreground extinction; previous 
estimates of the extinction have been mostly derived from the individual 
stellar colors; the derived extinctions are therefore not spatially continuous and their 
location along the line of sight is often uncertain. For both 
the HCN and P$\alpha$ observations, multiple fields were observed 
in order to image the entire Galactic center region out to and including the CND.

In \Section{2}, we describe the observations and reduction procedures. Three methods for removing the 
stellar continuum from the emission line images (scaled subtraction of the 190N image from the 187N image, PSF fitting and subtraction for the bright stars and variable scale factor image subtraction) are discussed in Appendix 1. In \Section{3}, we derive 
the extinction distribution  from the observed ratio
of P$\alpha$ to radio continuum and radio recombination lines and 
present an extinction-corrected P$\alpha$ image. Estimates of   
the sizes and densities of the HII structures are derived in \Section{4}, 
and the stellar sources with emission line excesses
are discussed in \Section{5}. We then examine the observed and extinction-corrected diffuse light distribution of stars in \Section{6} and the 
stellar density profile relative to \astar.

\section{Observations and Data Analysis}

\subsection{NICMOS P$\alpha$ Observations}

Images of the CND were obtained with the NICMOS camera on HST in the 
narrow-band 187N (P$\alpha$)  and 190N filters as part of the NICMOS 
GTO program.  A mosaic image, extending out to 50\arcsec\ radius, was 
constructed from four Camera 3 (NIC3) dithered fields and one Camera 2
(NIC2) dithered field (for higher resolution
near the center). NIC3 uses a 256$\times$256 HgCdTe array with plate scale of 
0.203\arcsec\ per pixel providing a 52.0\arcsec\ field of view; Camera 2 has 256$\times$256 0.076\arcsec\ pixels and a 19.3\arcsec\ FOV (Thompson \etal 1998).  The F187N and F190N filters with effective wavelengths of 1.87 and 1.90 $\mu$m were used to obtain on-line and off-line images. The F187N filter has a bandwidth of $\sim$ 1\% corresponding 
to over 3000\kms; it therefore includes all velocities of the 
Galactic center ionized gas (Lacy \etal 1991). The filter also includes a HeI(4-3) doublet line
(at 1.8691 and 1.8702 $\mu$m) ; for the extended emission, P$\alpha$ is expected to completely dominate (by more than a factor of 30), but on some emission lines stars (\cite{pau01}) the HeI line could be a significant
contribution. 

For each of the field centers (listed in Table 1), 4 dithers were taken spaced by 1.1\arcsec with
non-destructive reads at each position (MULTIACCUM).   The total
integration times for each filter are listed in Table 1. 
The diffraction limited resolution of HST is 0.19\arcsec\ at 1.87$\mu$m; the NIC2 images are therefore sampled at better than the Nyquist frequency but the NIC3 images are undersampled
by the Nyquist criterium. Although not completely recovering the 
Nyquist sampling, the half-pixel dithers do help to reduce the 
undersampling in NIC3.

\subsection{Data Reduction}

 The data were calibrated using the CALNICA version 3.3
task (\cite{bus98}) in IRAF\footnote{IRAF is distributed by 
the National Optical Astronomy Observatories, which are 
operated by the Association of Universities for Research 
in Astronomy, Inc., under cooperative agreement with the 
National Science Foundation.}/STSDAS with most reference
files (static data quality, detector read noise, detector
non-linearities files) from the Space Telescope Science Institute
(STScI) NICMOS pipeline. The flat-field and
dark frames were provided by the NICMOS Instrument Development Team (IDT). 
The dithered images were then shifted and coadded; for the NIC2 fields, using the 
the IDP3 routine from the NICMOS IDT software.  We estimate the registration
accuracy between the NIC2 fields to be better than 0.05 NIC2 pixel. The registration
accuracy for the NIC3 fields was less accurate due to the undersampling
of the point spread function (PSF) and was of order 0.25 NIC3 pixel (0.05\arcsec\ ).
The NIC3 data were dithered by half-pixel steps to oversample the original
pixels by a factor of 2 in the final mosaic. The NIC2 and NIC3 mosaics  were rotated to north up and east  to the left using the data-header orientation angle and uniformly 
sampled to 0.0753\arcsec\ square  pixels, corresponding to 0.0029 
pc or 9.00 $\times$ 10$^{15}$ cm at a distance of 8.0 kpc. (The 
pixels are factors of 1.00906 (NIC2) and 1.005 (NIC3) larger 
in x than in y -- this was corrected to make square pixels by interpolation).  
Where the NIC2 and NIC3 fields overlapped, offsets were determined from centroiding 
at least 20 common stars in the overlap regions.  In the central area covered by 
NIC2 with better sampling, the final mosaic was made from just the 
NIC2 pixels.  

For the NIC3 images, we employ scale factors of
 5.050$\times10^{-5}$ and  5.033$\times10^{-5} $
Jy~(ADU/sec)$^{-1}$ at  1.87 (P$\alpha$) and 1.90 $\mu$m, 
respectively  (\cite{rie03}).  For the NIC2 images, 
the scale factors are 4.086$\times10^{-5}$ and  4.330$\times10^{-5} $. 
The rms noise measured in the final NIC3 and NIC2 images in areas of 
very high extinction with relatively few stars is typically 3.9 and 
2.15 $\mu$Jy per pixel, respectively.

The mosaiced F187N and F190N images are shown in Fig.~\ref{187_190} with a logarithmic stretch on the grey-scale. In the F187N image diffuse emission (P$\alpha$) at the peak level of 
$\sim$ 200 $\mu$Jy per pixel may be seen; however, over 850 stars in the image
are at a level above this brightness. Also 
apparent in these images are several areas with far lower star counts than their
surroundings -- presumably due to small but very opaque dust clouds in the line of sight. Contours of HCN (1-0) emission at 3\arcsec\ resolution (\cite{chr03}) are superposed on the 1.87 $\mu$m image; several areas of decreased stellar
density are well correlated with molecular clouds seen in HCN (see Fig.~\ref{187_190}),
implying that the dense dust clouds are located near the Galactic Center 
as pointed out in an early analysis of this work in Stolovy \etal 2001.

\subsection{Continuum Subtraction and Emission Line Flux Calibration}

In order to image the P$\alpha$ line emission, it is necessary to remove the much stronger stellar point sources.  The removal of the stellar continuum from the 
F187N images is difficult since the stellar density is high, the intrinsic 
colors of the stars vary and most importantly, the reddening is highly variable
across the field. We experimented with three techniques for continuum 
subtraction (subtraction of fixed-scaled version of the F190N image, point 
source removal of the stars, and subtraction of a variable-scaled version of 
F190N) and the results are discussed in the Appendix to this article.
The final results presented below are based on the last technique.
A list of emission-line stars is provided in one of the sections below.

To convert the flux densities in the F187N continuum-subtracted image into an emission line flux, we adopted an effective FWHM for the filter of 190\AA~
or $\Delta \nu$ = 1.63 $\times$ 10$^{12}$ Hz.  
 
\section{P$\alpha$ Images}

Fig.~\ref{pa_full} shows the P$\alpha$ emission for the entire 
region imaged. The reference position of \astar~ ( $\alpha_{2000}$ = $17^{\rm h} 45^{\rm m}$ 40$^s.0383\pm0.0007$, $\delta_{2000}$ = $-29^\circ 00^\prime
28.069\pm0.014^{\prime\prime}$;\cite{yus99}) was located on the NICMOS image using the 
offsets from maser source IRS 7 given in Menten \etal (1997). The relative offsets of all other maser sources specified by Menten \etal (1997) (IRS 11NE, IRS 7, IRS 15 NE, IRS 10 EE, IRS 16 NW, IRS 16 NE and IRS 15NE) agreed to within one 0.0753\arcsec\~ pixel in the final mosaic image. Subtraction of a scaled F190N mosaic from the F187N mosaic should in principle yield an
image of the Pa$\alpha$ emission, but, as shown in Fig. ~\ref{187_091x190},  strong positive 
and negative stellar residuals (due to varying intrinsic colors, varying extinction, and 
photospheric P$\alpha$ emission and absorption) are apparent.  We instead removed the
continuum with a scale factor that varied spatially across the field to compensate for these issues as described in the Appendix.  This variable spatial scale factor continuum removal method was employed in the subsequent P$\alpha$ figures and anlaysis presented here.

The most prominent large-scale emission features in Fig.~\ref{pa_full} correspond to : 
the East-West 'bar' structure approximately 7\arcsec\ south of \astar,
the Northern arm which extends from -2\arcsec\ to over 25\arcsec\ north of \astar,
and the western arm approximately 14\arcsec\ west of \astar. With the 
high resolution and sensitivity of the P$\alpha$ images, these large-scale
features break up into numerous filaments/arcs which are presumably 
density enhancements in the ionized gas or individual 
ionization fronts at the edges of neutral clouds. Smaller 
scale emission peaks are seen on the 26 emission line stars.
The 'mini-cavity' appears as a clearly evacuated spherical hole in the emission
of diameter 2.7\arcsec\ at 1.7\arcsec\ W, 2.8\arcsec\ S  of \astar ~(\cite{zad90}) and is
as shown in Fig.~\ref{pa_full}. A weak P$\alpha$ emission line star appears at the center of this cavity (Stolovy \etal 1999). Another weak emission line star is seen at the center of the "eastern cavity" (4.6\arcsec\ E, 3.4\arcsec\ S ).

\subsection{Extinctions from P$\alpha$, Radio Continuum and Radio Recombination Lines}

The free-free and recombination line emission at radio wavelengths suffer negligible extinction. The observed ratios of P$\alpha$-to-radio free-free (and recombination line flux) can therefore be 
used to estimate the dust extinction -- both the line and free-free emission depend of the line-of-sight integrated emission measure, but only the P$\alpha$ line emission is significantly 
extincted by dust. 
   
For case B recombination, 
the intrinsic P$\alpha$ flux in a pixel is given by 
 
 $$ F_{P\alpha}  = 6.41\times10^{-26}\left({T_e \over 6000K } \right)^{-0.87} {n_e n_p l a_{pix} \over 4 \pi d^2 }~{\rm erg~s^{-1}~cm^{-2}}  \eqno (1) $$

\noindent where T$_e$, n$_e$ and n$_p$ are the temperature and volume densities of electron and protons, l is the path length in the ionized gas, a$_{pix}$ is the projected area of a pixel and d is the distance (Table 4.4 in \cite{ost89}; the P$\alpha$ emissivity used to derive the constant in Eq. 1 is appropriate to 
n$_e$ = 10$^4$ cm$^{-3}$ but the dependence on n$_e$ is very weak.) In Eq. 1 and the following, we will adopt T$_e$ = 6000 K (see Morris \& Yusef-Zadeh 1989; Simpson \etal 1997). Similarly, the radio free-free continuum flux density from a pixel is given by
 
 $$ S_{ff}  = 4.17\times10^{-13}\left({T_e \over 6000K } \right)^{-0.35} \left({\nu \over 5~Ghz } \right)^{-0.1} {n_e n_p l a_{pix} \over 4 \pi d^2 }~{\rm mJy}  \eqno (2) $$

\noindent where $\nu$ is the frequency of the radio emission. Thus the intrinsic 
P$\alpha$/free-free ratio is :

 $$ \left({F_{P\alpha} \over S_{ff}}\right)_{intrinsic}  = 1.54\times10^{-13}\left({T_e \over 6000K } \right)^{-0.52} \left({\nu \over 5 Ghz } \right)^{0.1} ~{\rm erg~s^{-1}~cm^{-2}~mJy^{-1}}.  \eqno (3) $$

Similar to the radio free-free continuum emission, radio recombination line emission also traces the same ionized gas as P$\alpha$ but suffers  
negligible extinction, providing an alternative means of estimating the extinction. Roberts \& Goss (1993) observed the H92$\alpha$ recombination line
from Sgr A West. If the line is emitted under LTE conditions 
(\cite{rob93}), then the line flux integrated 
over the velocity width of emission per pixel is given by 

  $$ S_{H92\alpha} \Delta v  = 1.29\times10^{-12}\left({T_e \over 6000K } \right)^{-3/2} {\nu \over 8.3~Ghz } {n_e n_p l a_{pix} \over 4 \pi d^2 }~{\rm mJy~\kms }  \eqno (4) $$

\noindent using Eq. 13.28 from Rohlfs \& Wilson (1996). The intrinsic 
P$\alpha$/H92$\alpha$ ratio is therefore :

 $$ \left({F_{P\alpha}} \over S_{H92\alpha} \Delta v \right)_{intrinsic}  = 4.977\times10^{-14}\left({T_e \over 6000K } \right)^{0.63} \left({\nu \over 8.3 Ghz } \right)^{-1} ~{\rm erg~s^{-1}~cm^{-2}~(mJy~\kms)^{-1}}.  \eqno (5) $$

\noindent The observed flux ratio can be less than that in Eq. 3 and 5 due to extinction 
at $\lambda$ = 1.87$\mu$m (P$\alpha$). Rieke (1999) derived A$_{1.60\mu m} $/A$_V$ = 0.176 and A$_{2.22\mu m} $/A$_V$ = 0.105 from NICMOS 
observations of stars in the Galactic center; interpolating between 
these values with a $\lambda ^{-1.6}$ power-law yields A$_V$ = 
7.24 $\times$ A$_{P\alpha}$. (This is nearly equal to the value obtained from 
 Rieke \& Lebofsky 1985.) We therefore use the relation

  $$ A_{V}  = 18.1\times log  \left( {{F_{P\alpha} / S_{ff H92\alpha}}_{intrinsic} \over {F_{P\alpha} / S_{ff H92\alpha}}_{observed}} 
\right)~{mag} . \eqno (6) $$

\noindent to estimate the visual extinction where S$_{ff H92\alpha}$ is the 
free-free or H92$\alpha$ flux. 

To estimate the extinction of the P$\alpha$ emission, we use the $\lambda$ 
= 6 cm radio map of Yusef-Zadeh \& Wardle (1993). This VLA radio image at 0.40\arcsec\ $\times $ 0.67\arcsec\ resolution has good sampling at low spatial frequencies;
the image therefore contains nearly all the emission on scales corresponding to the CND P$\alpha$ emission. 
The P$\alpha$ image was 
smoothed to the resolution of the radio image before computing the observed ratios of P$\alpha$ to radio
continuum.    

Fig.~\ref{av} shows the extinction distribution obtained from the observed ratios of P$\alpha$ to radio continuum using Eq. 6 and a histogram of the derived extinctions is shown in Fig.~\ref{av_hist}. The extinctions range 
from A$_V$ $\sim$ 20 to 50 mag and the median pixel extinction is A$_V = $ 31.1 mag. The median extinction is in good agreement with the median derived from the 
colors of the Galactic center stars (\cite{blu96}, \cite{cot00}). However, the extinction
exhibits large-scale gradients across the region (see Fig.~\ref{av}) and low extinctions (20-25 mag) are seen in front of the East-West bar and 
the Northern arm and in the direction of the IRS 16 cluster. The generally
low extinctions seen along the NE-SW swath south of \astar ~is  
consistent with the low extinction values derived for the stellar continuum 
in IRS 16 (\cite{blu96}). High extinctions (35-45 mag) are seen in the Western arm and in general along the outskirts of the P$\alpha$ emission.
This may be due to the increased extinction associated with dust in
the molecular gas along the periphery of the ionized region. 
The apparently low extinctions toward the P$\alpha$ emission line stars in the
central parsec are contaminated by photospheric excess P$\alpha$ and should be 
disregarded in interpreting the line-of-sight extinction in the diffuse gas.
The apparently high extinctions shown on \astar~ and a few arcsec to the north 
are spurious, due to the presense of non-thermal radio emission there; the
A$_V$ derived from the H92$\alpha$ line shows no anomaly on \astar~ (see below). 

The extinction derived from the 6-cm radio continuum image suffers
from two major uncertainties : non-thermal emission and the removal of spatially smooth, low-level
emission in the VLA  maps. The non-thermal contribution is significant on \astar  ~ and a small contribution could arise from Sgr A East (\cite{ped89}). For this reason we also derived the extinction using  
the H92$\alpha$ radio recombination line image obtained by Roberts \&
Goss (1993). This H92$\alpha$ data, smoothed to 2\arcsec\ resolution, has a considerably 
lower SNR (and spatial resolution) than the 6-cm image, but it is 
completely immune from problems of non-thermal emission contamination by \astar.
Fig.~\ref{av_h92} shows the extinction distribution obtained from the observed ratios of P$\alpha$ to H92$\alpha$ emission using Eq. 6. The 
areas with sufficient signal-to-noise ratio exhibit extinction values
in close agreement with those calculated from the 6-cm
data (see Fig.~\ref{av}). Using the H92$\alpha$ data, the median pixel
extinction is A$_V = $ 29.2 mag, compared with 31.1 mag from the 6-cm analysis
(see above). In Fig.~\ref{av_ratio} we show a histogram comparing the 
extinction extimates from H92$\alpha$ and 6-cm radio continuum 
on a pixel by pixel basis, illustrating the general consistency of the 
two approaches.  Because of the high sensitivity, resolution and spatial
coverage, we will use the extinction derived from 6-cm radio continuum in the
discussion below, but the agreement between the two methods  
reassures us that there are not large systematic errors. In addition,
the fact that both determinations exhibit a minimum in the ionized
bar region and Northern arm argues strongly that this minimum 
in the extinction is real. Lastly, we note that we also used a 
2-cm radio map (\cite{zad90}) for a parallel extinction analysis and derived very similar
estimates but this data has lower SNR than the 6-cm data. 

\subsection{Extinction-corrected P$\alpha$ and Lyman Continuum Luminosity}

The distribution of P$\alpha$ emission corrected for extinction at 1.87$\mu$m
is shown in Fig.~\ref{pa_unext}. Although the original radio continuum 
image is also, of course, extinction-free, the extinction-corrected P$\alpha$ image shown in 
Fig.~\ref{pa_unext} has the advantage of higher spatial resolution 
and signal-to-noise ratio. Under the assumption that the extinction
does not vary on scales smaller than the radio image resolution  
(0.40\arcsec$\times$0.67\arcsec), this image should provide the most accurate representation of the diffuse HII distribution. A histogram of the derived extinction-corrected P$\alpha$ fluxes (per pixel) is shown in Fig.~\ref{pa_hist}.

The total flux (within 40\arcsec\ radius of \astar) in the extinction-corrected P$\alpha$ image is
2.4$\times 10^{-9}$ ergs cm$^{-2}$ sec$^{-1}$; the luminosity is 
therefore 1.8$\times 10^{37}$ ergs sec$^{-1}$ or 1.7$\times 10^{49}$ P$\alpha$
photons per sec. For Case B recombination
at 6000 K (\cite{rob93}; \cite{rob96}), $\alpha_B(H)$/$\alpha_B(P\alpha)$ = 22.7 (\cite{ost89}) and the 
total Lyman continuum (LyC) emission rate required for the P$\alpha$ emission is 
3.94$\times 10^{50}$ sec$^{-1}$ or 2.70$\times 10^6$ \lsun (assuming 1.2 Rydbergs per Lyman continuum photon). Within 20\arcsec\ radius of \astar,
the P$\alpha$ flux is 50\% of that given above; thus 50\% of the ionizing 
photons are absorbed within 20\arcsec\ or 0.8 pc. All of the emission line stars are within this region. The
derived LyC production rates are {\it minimum} estimates since we assume Case B recombination with no Lyman 
continuum escaping the region. Given the apparent geometry of a tilted 
ring in the neutral gas, it is, in fact, very likely that the actual production rate
could be several times higher. The 8 emission line stars studied 
by Najarro \etal (1997) have a combined Lyman continuum emission rate of 3$\times
10^{50}$ sec$^{-1}$ and thus they can probably account for the 
ionization without any extra input from \astar. \astar~ has not yet been 
unambiguously identified as an infrared point source in either the continuum or
lines. Extended emission is seen in our images at the location of \astar~  but 
no discrete source is seen.  The Lyman continuum emission rate obtained 
from P$\alpha$ is consistent with that derived directly from
the radio continuum ($\sim 3\times 10^{50}$ sec$^{-1}$, \cite{gen94}).

\section{Physical Conditions in the Ionized Gas}

\subsection{Sizes of Ionized Structures}

The structures seen in the ionized gas are well resolved in the NICMOS
P$\alpha$ images. The emission filaments in the Northern arm and bar have typical 
FWHM $\sim$ 0.3 -- 1\arcsec\ (or 0.4 -- 1.2$\times 10^{17}$ cm$^{-3}$)
in their small dimension and up to 10 arcsec in their 
long dimension. The highly elongated, fine structure in the ionized
gas is strongly suggestive that the structures are often ionization 
fronts on the edge of more extensive neutral gas clouds or arms. The ionized shell at the 'mini-cavity' has a diameter 
of 2.7\arcsec\ or 3.2 $\times 10^{17}$ cm. Stolovy \etal (1999) discuss
the structure of the 'mini-cavity' in P$\alpha$ and [FeII] emission lines and 
identify the central star which they propose to be the source of the cavity. The age of the 
minicavity is only 500 yrs for assumed constant expansion velocities of 100  \kms. It
is possibly even a younger structure;  Zhao \& Goss (1998)
estimated  a current expansion rate of 200 \kms~ from radio proper motion measurements.

\subsection{Electron Densities (n$_e$)}

Since the extinction-corrected P$\alpha$ emission
is proportional to the volume-integrated emission measure, the electron density in the ionized gas can be obtained from the extinction-corrected P$\alpha$ image. Inverting Eq. 1, we find  :

$$ <n_e>~~ =~~1.21\times10^4~F_{P\alpha-14}^{1/2}~~T_{6000}^{0.44}~{D_{8} \over \left(l_{17} ~a_{pix}\right)^{1/2}}   
 ~~ cm^{-3}  \eqno (7) $$

\noindent where F$_{P\alpha-14}$ is the line flux in 10$^{-14}$erg~s$^{-1}$~cm$^{-2}$, T$_{6000}$ is the temperature normalized to 6000 K (\cite{rob93}, \cite{rob96}), D$_{8}$ in the 
distance to the Galactic center (8.0 kpc), l$_{17}$ is the line-of-sight pathlength (normalized to 10$^{17}$ cm corresponding to 0.83\arcsec\ ), and a$_{pix}$ is the projected area of the pixel normalized to 0.0753$\times$0.0753\arcsec\ . 
The peak extinction corrected P$\alpha$ fluxes along the Northern arm ridge are 4.3$\times 10^{-14}$ erg~s$^{-1}$~cm$^{-2}$  per
 pixel. Adopting $l_{17}$ = 1, the electron 
density in these peaks is n$_e$ = 2.5$\times 10^4$ cm$^{-3}$ and the 
mean n$_e$ along the Northern arm is n$_e$ $\sim 10^4$ cm$^{-3}$. These estimates are in good agreement with those originally derived by Lo \& Claussen (1983) from the radio free-free emission.

\subsection{Relative Geometry of Ionized and Molecular Features}

Some of the large-scale ionized structures can be ascribed 
to photoionization of the exterior molecular gas by Lyman continuum from 
hot stars in the central star cluster. The clearest example of this
is the Western arm. It extends over 70\arcsec\ (N-S)
in both the ionized and molecular gas tracers and the P$\alpha$ emission centroid is everywhere offset 5 -- 10\arcsec\ inside the HCN (Fig.~\ref{pa_unext}). In the western arm, the ionized and molecular gas 
velocities are also similar (\cite{chr03}). On the other hand, the ionized structures at smaller radii (e.g. the bar and lower Northern arm) do not have any obvious molecular counterparts. These 
may be structures which are almost entirely ionized, elongated dynamically and compressed laterally by winds (see below). The north end of the Northern arm in P$\alpha$ apparently connects smoothly in space and velocity (\cite{lac91}) to a molecular feature extending south ($\Delta \alpha$ = 10\arcsec\ , $\Delta \delta$ = 30\arcsec\ ). In this instance, possibly the neutral gas becomes ionized at $\Delta \delta \sim$ 18\arcsec\ as it approaches the central cluster. 
Jackson \etal (1993) have suggested
that the Northern arm is the ionization front on the surface of 
an atomic cloud seen in the 63$\mu$m [OI] line, which fills the interior 
of the CND but displays a twin peak morphology similar to that seen in
the integrated 2.12$\mu$m H$_2$ emission line (\cite{yus01}) associated
with the molecular torus of the CND.  Further evidence for a neutral component of
the gas interior to the CND was shown by Stolovy (1997)  from  [SiII]  34.8 $\mu$m 
observations at 15 \arcsec\ resolution, which peaked near the ''minicavity'' possibly as a result
of enhanced Si abundance due to grain destruction in shocks.

 Correlation (or lack of correlation) between the molecular gas mapped 
by the HCN contours in Fig.~\ref{av} and the derived extinctions in front of 
the ionized gas might be used to infer the relative placement of the 
H$_2$ and HII gas along the line of sight. The strong increase in 
extinction at the edge of the Western arm suggests that 
some of the ionized gas is behind, or mixed with, the neutral gas and dust
in the molecular Western arm. Increased extinction in front of the ionized gas is also apparent where two filaments of molecular gas (traced by the HCN contours in Fig.~\ref{av}) extend southward into the northern boundary of the ionized gas ($\Delta \alpha$ = -5\arcsec\ , $\Delta \delta$ = 15\arcsec\ , see Fig.~\ref{av}). 

However, Christopher \etal (2003) estimate volume densities n$_{H_2} \geq 10^6$ cm$^{-3}$ for the HCN molecular
clouds. Their typical sizes ($\sim 5$\arcsec\ or 7$\times 10^{17}$ cm) then 
imply column densities $\sim 7\times10^{23}$ H$_2$ cm$^{-2}$ or A$_V \sim$ 700 mag for 
a standard Galactic dust-to-gas ratio. Clearly, the extinctions in front of the 
the Western arm and northern P$\alpha$ emission are not nearly this high; if they
were, no emission would have been detected in the near-infrared. We therefore 
conclude that P$\alpha$ emission on the west and northern border of the region
($\Delta \alpha$ = -10\arcsec\ , $\Delta \delta$ = 10\arcsec\ ) must be on the front face of the molecular gas (unless the dust is 
severely depleted or the molecular gas is extremely clumpy on scales 
$<<$ 1\arcsec\ ).     

\subsection{Tidal Disruption and Stability}

The extraordinary small scale structures now seen in both the molecular and ionized gas in the central parsecs, must be extremely short-lived and evolving rapidly. The dynamical time (R/V$_{rot}$) is only 1.3$\times 10^4$ yrs
at R = 1.5 pc with V$_{rot}$ = 110 \kms ~(\cite{jac93}). If some of the 
ionized gas is infalling at $\sim$1/2 of the rotational velocity as
suggested by their shapes (for example the Northern arm as modelled by 
Vollmer \& Duschl 2001 and Sanders 1998), then the infall times
are only a few 10$^4$ yrs. 

Similar short timescales are derived 
from the small scale structure in the P$\alpha$ emission. Along the Northern arm, the width of the bright emission ridge is just 10$^{17}$ cm (corresponding to FWHM = 0.83\arcsec~); if this gas freely expands at the 10 \kms~sound
speed of the HII, the time required to double its width is just
3$\times 10^3$ yrs. Thus if the ionized gas is to maintain its high
surface brightness over a dynamical or infall time, the observed 
emission ridges must be ionization fronts on the edge of dense,
neutral material or the ionized gas must be compressed by a much lower density
and higher temperature confining medium (e.g. a wind from \astar~or the 
mass-loss stars -- see below) or magnetic fields. 
Most likely,  both scenarios are occurring. 
The ionized gas on the Western arm is well-correlated both spatially 
and kinematically with the dense molecular gas seen in HCN emission 
(\cite{chr03}) and thus is likely to be the ionized inner face of the 
molecular arm. 

A number of investigations starting with Quinn \& Sussman (1985) have emphasized the importance of tidal disruption to the central parsec gas clouds. The Roche limit associated with the central 2.6$\times 10^6$ \msun ~
(i.e. the \astar ~black hole; \cite{gen00}; \cite{ghe00}) is
n$_H \geq 2.4\times 10^8$/R$_{pc}^{3}$~ cm$^{-3}$ (e.g. \cite{san98}). This 
is a factor of 10$^4$ (!) greater than the density estimated above 
for the ionized gas in the Northern arm and is comparable to or greater than 
the densities derived by Christopher \etal (2003) for the HCN clumps at 
$\sim 2$ pc radius. Inclusion of the mass associated with the stellar distribution
doubles the Roche limit  at the radius of the CND.

The data clearly show the ionized gas in the Northern arm directly connecting to a N-S feature in the molecular gas. The kinematics of the [NeII] (\cite{lac91}) and HCN emission are also 
consistent with this connection.  Thus, the Northern arm is 
very likely an infalling streamer originating from the 
northern molecular gas (Jackson \etal 1993, Wright \etal 2000). The molecular gas might have been tidally sheared  and ionized as it approaches \astar ~and the hot central
stars. 

The sharpness of the Northern arm in cross-section is illustrated 
in Fig.~\ref{slice} where a 1-pixel wide slice in the E-W direction is shown for the 
observed and extinction-corrected P$\alpha$. The width is less than 1\arcsec\ . Confinement of the ionized gas in the east-west direction
requires a confining medium, magnetic field, or wind ram pressure. Paumard \etal (2001) show that there are two classes of emission line stars :
those with broad emission lines ($\Delta V \sim 1000$ \kms) ~and those 
like the IRS 16 sources with widths of $\sim 200$ \kms. The mass-loss 
stars in the central He cluster have a total $\dot{M}$$\sim 3\times 10^{-3}$ \msun ~yr$^{-1}$ and typical velocities v$_w \simeq 750$ \kms (\cite{hal82}; \cite{kra91}; \cite{naj97}). 
At a distance of 10\arcsec\ (0.4 pc) from the mass-loss stars, the wind will
result in a maximum ram pressure $\rho v^2$$\sim 9\times10^{-7}$ dynes cm$^{-2}$
(assuming no deceleration of the wind). The thermal pressure ($nkT$) of the ionized 
gas in the Northern arm is approximately two orders of magnitude less
(8$\times10^{-9}$ dynes cm$^{-2}$ assuming n = 10$^4$ cm$^{-3}$ and
6000 K). Thus, the mass-loss winds might reasonably contribute to 
the confinement of the ionized gas in the arm.   
It is also noteworthy that Baganoff \etal (2001) have recently detected 
X-ray emission from hot, ionized gas in the vicinity of \astar. This gas could  drive an outflowing wind, which might also compress the HII gas in the Northern arm.    
Aitken et al. (1991) concluded
from mid-infrared polarimetry that the magnetic field is aligned parallel to the flow
of ionized material along the filaments and that the magnetic field
is therefore magnified due to these shearing motions. They also concluded
that the IRS16 sources do not disturb the magnetic field and therefore they
are likely out of the plane of the Northern arm.

Thus, it seems likely that the Northern arm gas must be confined in 
the E-W direction by a wind or a confining medium from the general direction of \astar
(but not necessarily from the IRS 16 sources--the IRS13 wind sources and \astar
itself may contribute to the wind).
The shape of the Northern arm, concave around \astar, is also suggestive of 
pressure confinement on its west side. In the east direction (away from \astar)
the P$\alpha$ emission in the Northern arm falls off more gradually and hence does not require confinement. 

\section{Candidate P$\alpha$  Emission Line Stars}

We used the IDL 'starfinder' (Diolaiti \etal 2000) program to derive the fluxes in each of the F187N and F190N filters by PSF matching for the NIC2 data only, which led to satisfactory results in deriving stellar fluxes as compared to aperture photometry and scaled PSF subtraction.
(The NIC3 portion of the mosaics, despite dithering and resampling of the pixels, did not
have not sufficiently well defined PSF's to use PSF-matching techniques).  
We then correlated well-matched sources found in both filters and computed the F187N/F190N ratios. Although the large width (1\%) of the NICMOS F187N filter makes it difficult
to detect stars with relatively weak  P$\alpha$ line emission on top of the 
strong stellar continuum, we do list 26 stars 'candidate' emission line stars in Table 2 
which were chosen based on derived F187N/F190N flux ratios $\geq$ 1.1.  
The typical  F187N/F190N ratio for stellar sources was found to be 
0.91; a ratio of 1.1 is therefore a conservative indicator of 
significant ($\sim 20\%$) excess flux at 1.87$\mu$m. 
The high reddening of the Galactic center stars implies F187N/F190N $<$ 1
for all stars.  The fluxes in Table 2 have not been corrected for extinction.

The last column in Table 2 lists the likely identification of the stars with previously observed HeI and HI emission line stars (\cite{gen96}; \cite{naj97}; \cite{pau01}).  Several of these candidate P$\alpha$ emission line stars are also strong HeI (2.058$\mu$m) emission line stars and the F187N flux may indeed be dominated by the HeI line within the bandpass rather than P$\alpha$.  Some stars in Table 2 appear to be newly identified.  
(We note that some of the stellar positions listed here disagree with those given by Paumard \etal (2001) by as much as 0.8\arcsec\ ; we have discussed this discrepancy with 
Paumard \etal and their revised positions are in better agreement.) We are confident of our positions since the reference stars given by Menten \etal (1997) agree to within 0.07\arcsec~
in the NIC2 image.  Several stars with high F187N/F190N ratios were removed from
the candidate list after visual inspection determined that the excess was dominated by diffuse interstellar P$\alpha$ emission along the line of sight.  Weaker P$\alpha$ emitters, such as IRS16CC,  the 'minicavity star',  and the  'eastern cavity star' do not appear in Table 2, as they do not obey the criterion set by F187N/F190N flux ratios $\geq$ 1.1.   'Starfinder' did not resolve the 2 emission line stars located near IRS13E, now called IRS13E2 and IRS13E4 (Maillard \etal 2003). 
The fluxes for these stars were derived by a scaled subtraction of the PSF from each
star in each filter.   IRS15NE appears as 2 stars: a P$\alpha$
absorption line star to the south coincident with the maser position of IRS 15NE and a star
0.5\arcsec to the north, which is an emission line star only imaged in NIC3. Aperture
photometry was used to derive the P$\alpha$ flux for the northern source, which we
call IRS 15NNE.

\section{Centroid and Radial Distribution of Stellar Continuum}

It is of interest to determine if \astar ~is located near the 
center of the nuclear star cluster and the ionized gas. Previous 
ground-based observations of the stellar light distributions 
have included the contribution from all stars, including the 
bright stars. Here we investigate the diffuse light surface brightness distribution 
by measuring the median pixel brightness as a function of radius from 
\astar and the mean of pixels excluding the 20\% of the pixels 
with highest brightness. The very clean and stable PSF of NICMOS on HST
is particularly advantageous for this analysis. 
We have also measured the centroids of the total P$\alpha$ and 1.9$\mu$m stellar continuum flux. Although
the starlight may not have the same effective foreground extinction
distribution as the ionized gas, it is likely the stellar light 
corrected for the extinction derived for the ionized gas is a much better 
approximation to the intrinsic 1.9$\mu$m continuum than the observed 
1.9$\mu$m continuum. For the extinction distribution we use that 
shown in Fig.~\ref{av} with the exception that within $\pm$ 0.5\arcsec\ of \astar, the extinction was set to the average between 0.5 and 0.75\arcsec\~ 
radius which is within 10\% of that determined from H92$\alpha$. The non-thermal flux of \astar ~yields a spuriously high apparent extinction on \astar. In computing the brightness distributions, we also exclude the very bright source IRS 7. At each radius in the distributions,
 we only include those pixels within the mosaic 
area with derived extinctions (and in the case of P$\alpha$ only those 
pixels with detected P$\alpha$).  

The emission centroids for the total P$\alpha$ and the 1.9$\mu$m flux 
were measured for apertures of radius 2, 5, 10, 15, 25, 30, 35, 40\arcsec\ ~from \astar. These centroids are plotted as boxes in Fig.~\ref{centroid} on top
of the 1.9$\mu$m image of the region within 2.5\arcsec\ of \astar. For both the
ionized gas and stellar continuum, the centroids remain within  
$\pm$ 2\arcsec\ over the entire range out to 40\arcsec\ ~radius. This provides 
very strong evidence that \astar (the putative central black hole) is
indeed at or extremely close ($\leq 0.04$ pc or $1.2\times10^{17}$ cm) to the center of the Galactic nucleus stellar distribution and presumably the 
dynamical center of the Galaxy. 

The radial distribution of observed and extinction-corrected 1.9 $\mu$m flux density is shown in Fig.~\ref{radial}. We
removed the very bright star IRS 7 and the bins in the distributions are independent. The greatest difference between 
the observed and extinction-corrected distributions occurs at radii
outside 10\arcsec\ where the larger extinction corrections associated
with the CND raise the extinction-corrected flux by a greater amount. The large scale 
near-infrared light distribution was originally measured by 
Becklin \& Neugebauer (1968) who fitted an R$^{1.2}$ power-law 
at radii between 6\arcsec\ and 10\arcmin\ . ~The extinction-corrected
distribution shown in Fig.~\ref{radial} is clearly somewhat steeper 
than R$^{1.2}$ and closer to the R$^{1.8\pm0.2}$ power-law shown beside the data. 
This flux distribution corresponds to a radial density distribution 
$\rho \propto R^{-0.5}$ in this outer region. This distribution is 
considerably flatter than that derived from number counts of 
individual stars ($\propto$ R$^{-2}$, \cite{gen96}, \cite{all94}, \cite{rie94})).
These steeper dispersions are strongly influenced by the 
'ring' of early-type and giant stars at radii of 4-10 \arcsec\ (\cite{gen96}).
 The outer radial flux distribution (corrected for extinction)
is fit by the power law but a noticeable excess flux is 
seen in the range 0.8 to 10\arcsec\ radii in Fig.~\ref{radial}. This excess is probably partially due to the young stars in the IRS 16 cluster. 

The mean surface brightness distributions of the {\it diffuse} starlight are shown in 
Fig.~\ref{sb}. Here, to avoid the strong effects of bright individual stars,
the surface brightness (both mean and median pixel brightness) were
computed from the pixels in the lower 80\% of the brightness
distribution at each radius. (The results are insensitive to the adopted 
80\% cutoff since cutoff between 50 and 85\% were tested with similar results;
the 80\% cutoff was adopted finally since it makes use of a large fraction of the pixels. The median also attenuates the effects 
of a few very bright pixels.) Both the observed and extinction-corrected
brightness distributions increase at smaller radii but exhibit 
a sharp drop at 0.8\arcsec\ radius, corresponding to 0.03 pc or 1.0$\times 10^{17}$ cm. To evaluate whether this drop is statistically 
significant, we independently evaluated the mean and medians
for the 4 quadrants (NE, SE, SW, NW) from \astar and then estimated
the uncertainties from the standard deviation of these independent samples 
-- these are the vertical error bars shown in Fig.~\ref{sb}. The central 
drop is significant given these uncertainty estimates.   

The density of discrete stars (as opposed to the total or diffuse light 
distribution) has been derived as a function of projected radius by 
Genzel \etal (1996). They also found a decrease in the projected 
surface density of star counts. For stars with K $\leq$ 12 mag, the drop 
occurs at 3\arcsec\ ~radius; for brighter star with K $\leq$ 10.5 mag, 
the decrease occurs inside 5\arcsec\ ~radius. Thus our results differ
significantly from theirs, showing the drop occuring at $\sim$ 0.8\arcsec\ .
Genzel \etal (1996) attribute the decrease at small radii as being due to  
collisions of the giant stars with main sequence stars which then deplete
the giant stars because of their relatively large cross-sections. 

To test whether 
the drop in the surface brightness distribution inside 1\arcsec\ is due to 
a lack of bright giant stars inside 0.8\arcsec\ , we also
show in Fig.~\ref{sb}, the median pixel surface brightness. This median flux 
should be virtually unaffected by the bright stars 
which dominate a relatively small fraction of the pixels; it still exhibits 
a modest decrease at small radii, although not as large as that seen in the 
mean surface brightness.

The drop or leveling off of surface brightness inside 1\arcsec\ could
have several explanations: it might be the core radius of the nuclear 
stellar distribution or it might be due to the depletion of 
late type stars with high mass-to-light ratios (Phinney 1989; Bailey \& Davies 1999). Gravitational lensing would produce a drop inside the Einstein 
radius which is on the scale 0.01\arcsec\ (\cite{war92}) and is therefore 
not a likely explanation. Sellgren \etal (1990) found a diminishing 
depth of the 2.3$\mu$m CO stellar absorption feature in the central few 
arcsecs, possibly indicating a depletion of giant stars at the center.
An alternative possibility is ejection of stars from the 
central 1\arcsec\ by a past or present binary black hole (Milosavljevic
\& Phinney -- private communication).

\section{Summary}

We have used a mosaic image of the 1.87$\mu$m HI P$\alpha$ line  to estimate the extinction distribution in the Galactic nucleus by comparison with the 6-cm radio continuum 
and to derive the spatial distribution of ionized gas in the 
central 40\arcsec\ at 0.2\arcsec\ resolution (Nyquist sampled in 
the central 19\arcsec\ ). 

1) In front of the ionized gas, A$_V$ varies from 20 to 50 mag with a median pixel value 
of A$_V = $ 31.1 mag, in excellent agreement with earlier estimates derived for the stellar sources.

2) A broad minimum in the extinction extends $\sim$ 30\arcsec\ NE--SW over the near the brightest ionized emission (Northern arm, IRS16 region, E-W Bar) and increases in the extinction are seen along the periphery of the ionized gas towards the dense CND molecular features.

3) The minimum Lyman continuum emission rate from within 40\arcsec\ radius of \astar ~is 3.9$\times 10^{50}$ sec$^{-1}$ or L$_{LyC} \simeq 2.7\times 10^6$ \lsun. At least half of this emission is within 20\arcsec\ radius.

4) Mass-loss winds from 
the hot emission line stars are probably responsible for containment of many
of the high brightness the ionized filaments.

5) For both the
ionized gas and stellar continuum, the centroids of the large scale emission (corrected for extinction derived from P$\alpha$) remain within $\sim$ 
$\pm$ 2\arcsec\ of \astar  from 2\arcsec\ out to 40\arcsec\ ~radius. This provides 
very strong evidence that \astar (the putative central black hole) is
indeed at or extremely close ($\leq 0.04$ pc or $1.2\times10^{17}$ cm) to the center of the Galactic nucleus stellar distribution and presumably the 
dynamical center of the Galaxy.

6) The extinction-corrected 
enclosed stellar continuum flux is fit quite well by a R$^{1.8}$ power-law
at large radii with an excess from 1 to 8\arcsec\ radius. The 1.9$\mu$m surface brightness increases
strongly into 1\arcsec\ ~radius and exhibits a drop with 0.8\arcsec\
~of \astar, possibly indicating the influence of the central 
black hole on the nearby stellar distribution.

\vspace{5mm}

This paper is based on observations with the
NASA/ESA Hubble Space Telescope obtained at the Space Telescope Science
Institute, which is operated by Association of Universities for Research
in Astronomy, Incorporated, under NASA contract NAS5-26555. The
NICMOS project was supported by NASA grant NAG 5-3042 to the NICMOS
instrument definition team. We thank Doug Roberts for use of 
his H92$\alpha$ image. It is a pleasure to thank Zara Scoville 
for assistance in editing. 

\section{Appendix- Stellar Continuum Subtraction}

Subtraction of the stellar continuum emission from the 1.87 $\mu$m image was one of the most challenging aspects of this study. 
Compounding the problem of continuum subtraction from the 
line+continuum image (F187N), the stellar sources
have different intrinsic colors and variable reddening. In addition, some stars have P$\alpha$ emission or absorption. 

For subtraction of the stellar continuum from the F187N image, we experimented with three techniques: fixed-scale factor subtraction of F190N from F187N;
PSF fitting and subtraction of the stars from the F187N image; spatially variable scale factor subtraction of F190N from F187N. We discuss these 
approaches separately in the following subsections.

\subsection{F187N - C$\times$F190}

The most straightforward approach is to multiply the F190N image by a fixed scale factor and subtract this from the F187N image, with the scale factor 
being that which yields the smallest positive and negative residuals 
at the locations of identified stars. From trials with the scale factor 
varying between 0.85 and 1.0 in steps of 0.01, we determined that the best overall scale factor was 0.91 and the resultant residual image is shown in 
Fig.~\ref{187_091x190}.  This scale factor is confirmed in the peak of the histogram
of F187N/F190N ratios shown in Fig. ~\ref{187_190ratios}. The bright 
positive and negative spots in the residual image are due to
too little or too much stellar continuum being subtracted for the 
individual stars. Positive residuals predominate in the central 
region, negative residuals near the periphery. This systematic variation
clearly indicates that the reddening is lowest in front of the central area and increases over the periphery.
Thus a straightforward
subtraction of the F190N image (multiplied by a fixed-scale factor)
from the F187N image leaves strong negative and positive residuals at the positions of the stars. This implies that suitable 
continuum subtraction requires different scale factors at the location 
of each individual star. 

\subsection{Stellar PSF Subtraction}

The second approach removed the stars directly via PSF subtraction.  For the
PSF fitting we employed the 'starfinder'  IDL software package (Diolaiti \etal 2000).
The software is designed for imaging with oversampled PSFs; thus, PSF fitting
was only attempted for the NIC2 mosaic, which only marginally meets this
criterion.  A composite PSF was derived from the NIC2 mosaic itself and was dominated
by the brightest source in the field, IRS7.  PSF subtraction was attempted both by
subtracting an image of the detected stars (detected fluxes convolved with the PSF) from
the F187N directly as well as subtracting stars detected at both F187N and F190N from
the fixed scale factor Pa$\alpha$ image. We found that PSF fitting with
'starfinder' worked satisfactorily for deriving stellar fluxes in the F187N and F190N images
which was used to construct a list  P$\alpha$ emission stars.
However, we found that the PSF subtraction of the reconstructed
point sources did not give as satisfactory results in removing stellar residuals,
possibly due to a non-ideal PSF derivation coupled with the only marginal sampling.
This undersampling means that pixelation errors which are quite unpredictable 
become important.  In addition, the strongly varying diffuse P$\alpha$ background
over both small and large scales makes the PSF fitting less accurate than if the field
were entirely composed of point sources. We therefore decided after much effort that  PSF fitting 
and subtraction would not yield acceptable results in removing the stellar continuum.

\subsection{F187N - C(x,y)$\times$F190}

Lastly, we experimented extensively with a procedure which determined 
individual scale factors for approximately 1000 stars identified in the F190N 
and F187N images, used these scale local factors in the vicinity
of each of the stars and then adopted the average stellar scale factor 
for all remaining pixels. This produced a final residual image, shown in Fig. ~\ref{pa_full}
considerably cleaner than that obtained using a fixed scale factor (Fig.~\ref{187_091x190}), although the procedure involved a number of complex steps.   

Specifically, :

1) On both the F187N and F190N images, we first removed the diffuse
low-level background (P$\alpha$ and faint stars) determined 
as the median intensity in the  
16 $\times$ 16 pixel area centered on each original 0.0753\arcsec\ pixel. (The value of 16 pixels was chosen to be enough that at least 50\% of the pixels would 
be outside the PSF of bright stars, yet small enough that variations in the 
background could be picked up.) 

2) The F190N pixels associated with each stellar peak were then collected 
using the algorithm described by \cite{sco01}. In order 
to be considered as a star, the peak pixel had to exceed 410 $\mu$Jy in the 
F190N image; and the adjacent pixels were included in the star
down to 175 $\mu$Jy. Thus the pixel boundaries of over 1500 stars were determined.

3) Local scale factors for {\it each} star were then derived as the ratio
of the total flux measured 
in the F187N and F190N images within identical pixel boundaries for each star defined from F190N. The derived scale factors for 1031 stars are 
shown in Fig.~\ref{187_190ratios}.

4) An image of the continuum subtraction scale factors (C(x,y)) was then made 
by setting the local scale factor within each star's pixel boundaries
to be that derived individually in step 3, and setting the scale 
factor for all pixels not included in any star to the average of 
all the stellar scale factors.\footnote{Any stars with apparent 
F187N/F190N  flux ratios greater than 1.1 --a 20\% effect since the peak
scale factor is 0.91--were deemed to probably
be strong P$\alpha$ emission line stars (see below); for these stars, the local 
scale factor was also set to the average scale factor of the remaining stars.}
The use of the 'average' scale factor in the 'non-stellar' pixels does not 
introduce serious errors -- the precise scale factor there makes little
difference since the amount of F190N signal which is to be subtracted
is lower than the Palpha in F187N in these areas.

5) A difference image was then constructed as F187N - C(x,y)$\times$F190N. 

6) The difference image was generally better in terms of 
lower residuals (compared to those in Fig.~\ref{187_091x190}), but 
still had small residuals near some of the brighter stars (due probably
to pixelation errors). Therefore, above the 410 $\mu$Jy in the stellar
positions, we interpolated from the immediately
surrounding pixels.\footnote{In the vicinity of the very bright 
star IRS 7 which had a peak continuum surface brightness $\sim$ 30,000 $\mu$Jy per pixel, the area with strong PSF effects 
(30$\times$25 pixels) was entirely 
removed and interpolated.}  

Although the above process was complex and the last step (\# 6) is 
obviously {\it ad hoc}, the resulting image retains the maximum areal coverage,
replacing only the brightest stellar pixels with interpolated values, and 
results in an image relatively free of residual stellar effects. The focus of the present paper is the interstellar ionized gas and extinction distributions. The ionized gas in the directions of the stars is clearly more accurately 
studied with high resolution spectroscopy.

\clearpage

\figcaption[f1a.epsi,f1b.epsi]{The mosaiced F187N and F190N images are shown with pixels sampled to 
0.0753\arcsec\ and rotated with north up. The region is dominated by stellar
sources, and the P$\alpha$ emission is seen as relatively faint diffuse 
emission in the F187N image. The color scale units are $\mu$Jy per 0.753\arcsec\ pixel and the offsets are relative to \astar ($\alpha_{2000}$ = 17$^h$45$^m$40$^s$.058, $\delta_{2000}$ = -29\deg00\arcmin27.90\arcsec). 
In the 1.87$\mu$m image, contours of HCN (1-0) emission are shown and the boxes indicate the positions
of the P$\alpha$ emission stars listed in Table 2. The HCN countours are dashed when negative (i.e. absorption)
and the levels are at : -100,-50,-25,-10,-5,2,4,6,8,10,15,20,25,30,35,40,45,50,60
Jy km/ sec /beam. 
\label{187_190}}

\figcaption[f2.epsi]{P$\alpha$ emission for the entire region imaged. The color scale units are $10^{-16}$ ergs cm$^{-2}$ sec$^{-1}$ per 0.0753\arcsec\ pixel and spatial
offsets are relative to \astar.
\label{pa_full}}

\figcaption[f3.epsi]{The extinction distribution derived from the observed
ratio of P$\alpha$ to 6-cm radio continuum emission. The P$\alpha$ emission 
was convolved to the same spatial resolution as the radio and we assume the 
radio continuum is entirely free-free except on \astar. The contours 
(-100,-50,-25,-10,-5,2,4,6,8,10,15,20,25,30,35,40,45,50, and 60 Jy \kms) are for the HCN emission at 3\arcsec\ resolution (\cite{chr03}). A broad
minimum in the extinction is seen in the IRS 16 cluster and the extinctions
increase at the periphery of the ionized gas in the molecular features. The dashed contours in the HCN indicate locations where the HCN appears in absorption of the radio free-free (from the ionized gas) and nonthermal 
(\astar) continuum. The {\it apparently} high extinction on \astar is 
due to the strong non-thermal contribution which yields spuriously 
large extinction estimates.  Likewise, the interstellar extinction towards the bright P$\alpha$ 
emission line stars cannot be accurately measured; the excess photospheric
P$\alpha$ emission yields spuriously low extinction values towards these stars.
\label{av}}

\figcaption[f4.epsi]{Histogram of visual extinctions (A$_V$)
for pixels with detected radio continuum and P$\alpha$ emission. The median extinction is A$_V$ = 31.1 mag. 
\label{av_hist}}

\figcaption[f5.epsi]{The extinction distribution derived from the observed
ratio of P$\alpha$ to H92$\alpha$ radio recombination line emission (\cite{rob93}). The P$\alpha$ emission 
was convolved to the same spatial resolution as the radio (2\arcsec\ ). The contours 
are for the HCN emission at 3\arcsec\ resolution (\cite{chr03}). This extinction distribution agrees closely with that derived independently 
from the 6-cm radio continuum (Fig.~\ref{av}). A broad
minimum in the extinction is seen in the E-W bar and Northern arm  and the extinctions
increase at the periphery of the ionized gas in the molecular features.  We note
that regions with insufficient signal/noise in the H92$\alpha$ image are masked out
(such as the region to the SW of Sgr A*).  The dashed contours in the HCN indicate locations where the HCN appears in absorption of the radio free-free (from the ionized gas) and nonthermal 
(\astar) continuum.
\label{av_h92}}

\figcaption[f6.epsi]{Histogram of the ratios of extinctions determined 
from P$\alpha$/H92$\alpha$ versus those determined 
from P$\alpha$/6-cm radio continuum. The histogram is of course 
restricted to those pixels shown in both Figs.~\ref{av_h92} and ~\ref{av}
with determinant extinctions.
\label{av_ratio}}

\figcaption[f7.epsi]{The P$\alpha$ emission corrected for extinction using 
the extinction derived from the ratio of P$\alpha$ to radio continuum emission (Fig.~\ref{av}). The color scale units are $10^{-16}$ ergs cm$^{-2}$ sec$^{-1}$ per 0.0753\arcsec\ pixel and spatial
offsets are relative to \astar. The contours 
are for the HCN emission at 3\arcsec\ resolution (\cite{chr03}). 
\label{pa_unext}}

\figcaption[f8.epsi]{Histogram of extinction corrected P$\alpha$
fluxes (per 0.0753\arcsec\ ~pixel)
for pixels with extinction estimates derived from the ratio of P$\alpha$ emission to radio continuum.  
\label{pa_hist}}

\figcaption[f9.epsi]{The P$\alpha$ intensity distribution along an 
E-W strip (1 pixel wide) crossing the Northern arm is shown.
The peak occurs at 7.2\arcsec E, 11\arcsec N of \astar . The two traces are for the 
observed (solid line) and extinction-corrected (dashed line) P$\alpha$ fluxes.  
\label{slice}}

\figcaption[f10.ps]{Position centroids of extinction-corrected P$\alpha$ (small boxes) and 
extinction-corrected 1.9 $\mu$m (large boxes) emissions are plotted on 
the observed 1.9 $\mu$m emission within $\pm$ 2.5\arcsec\ of \astar. The centroids 
were measured successively for pixels within 2, 5, 10, 15, 25, 30, 35, 40\arcsec\ ~radii of \astar ~and each box is labelled with the corresponding radius. The 
very bright source IRS 7 was excluded and the extinction within $\pm$ 0.5\arcsec\ of \astar ~was set equal to the average in the annulus between 0.5 and 0.75\arcsec\  
radius (on account of the non-thermal flux of \astar ~which precluded 
extinction estimates directly on \astar). 
\label{centroid}}

\figcaption[f11.epsi]{The radial distribution of observed and extinction-corrected 1.9 $\mu$m flux densities are shown as 
a function of aperture radius. The 
very bright source IRS 7 was excluded and the extinction within $\pm$ 0.5\arcsec\ of \astar ~was set equal to the average between 0.5 and 0.75\arcsec\  
radius (on account of the non-thermal flux of \astar ~which precluded 
extinction estimates directly on \astar). Also shown are an R$^{1.2}$
power-law which Becklin \& Neugebauer (1968) fit to the observed 2.2 $\mu$m 
flux distribution on larger angular scales (6\arcsec\ to 10\arcmin\ ).
The R$^{1.8}$ power-law fits somewhat better to the extinction-corrected 
1.9 $\mu$m flux distribution.
\label{radial}}

\figcaption[f12.epsi]{The surface brightness distribution for the observed and  extinction-corrected 1.9 $\mu$m radiation is shown. Discrete, 
bright sources (such as IRS 7) were excluded by removing the brightest 20\%
of the pixels in each radial bin. Because the noise is very non-Gaussian
(dominated by the stellar brightness distribution), the vertical 
error bars were estimated from the standard deviation of values calculated 
in the 4 quadrants relative to \astar . 
\label{sb}}

\figcaption[f13.epsi]{The image of P$\alpha$ emission obtained by 
subtraction of 0.91$\times$F190N from the F187N image. The bright 
positive and negative spots in the residual image are due to
too little or too much stellar continuum being subtracted for the 
individual stars. Note that positive residuals are prevalent in the central 
regions and negative residuals near the periphery -- a result of the fact that
the reddening is lowest in front of the central area and increases
in the periphery.  In addition, the P$\alpha$ emission line stars (see Table 2)
are concentrated in the central region.
\label{187_091x190}}

\figcaption[f14.epsi]{The distribution of measured stellar flux ratios
F187N/F190N is shown for 1031 stars determined with the spatially variable scale
factor method described in the Appendix.
\label{187_190ratios}}

\newpage
\plotone{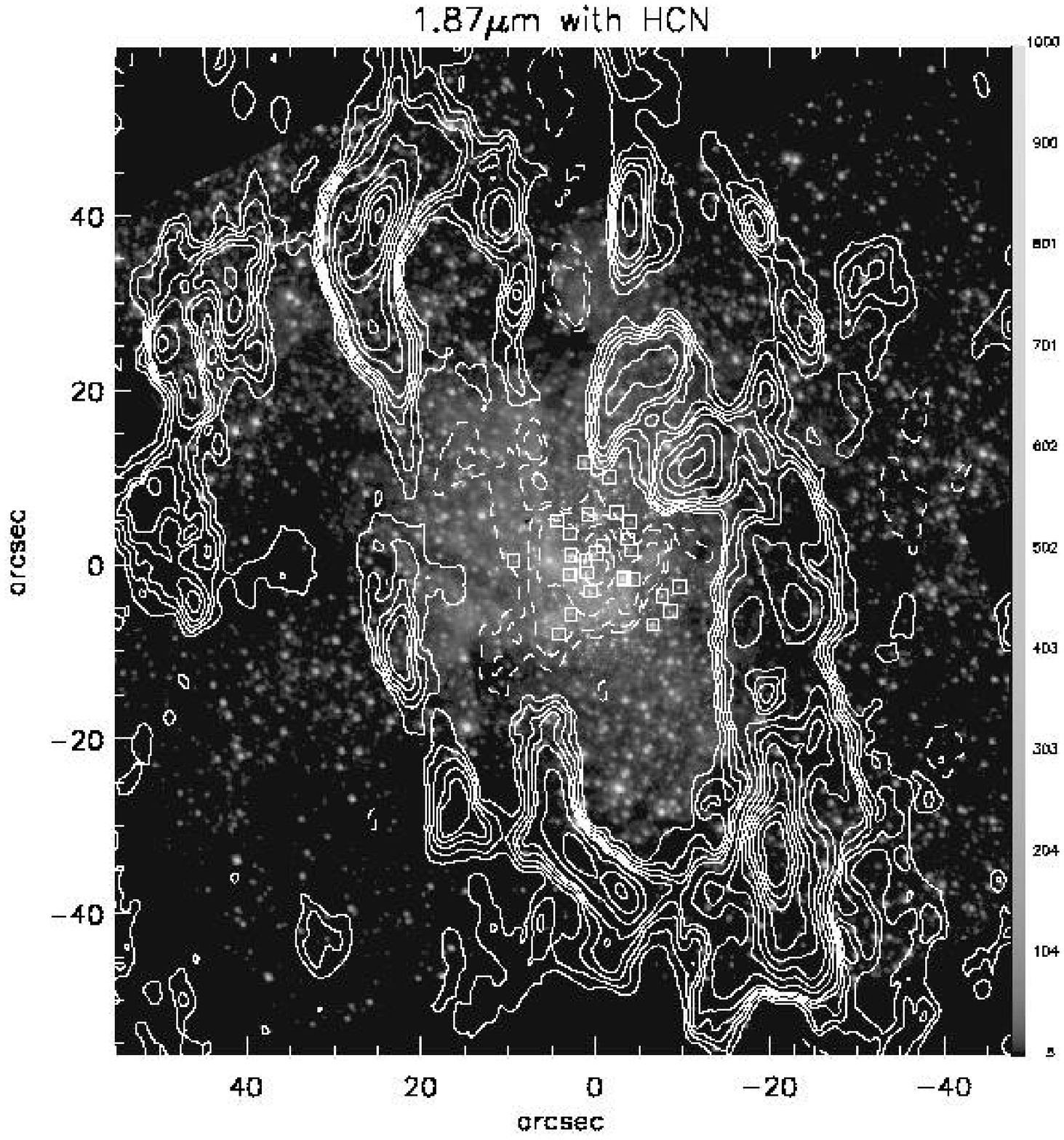}
\newpage
\plotone{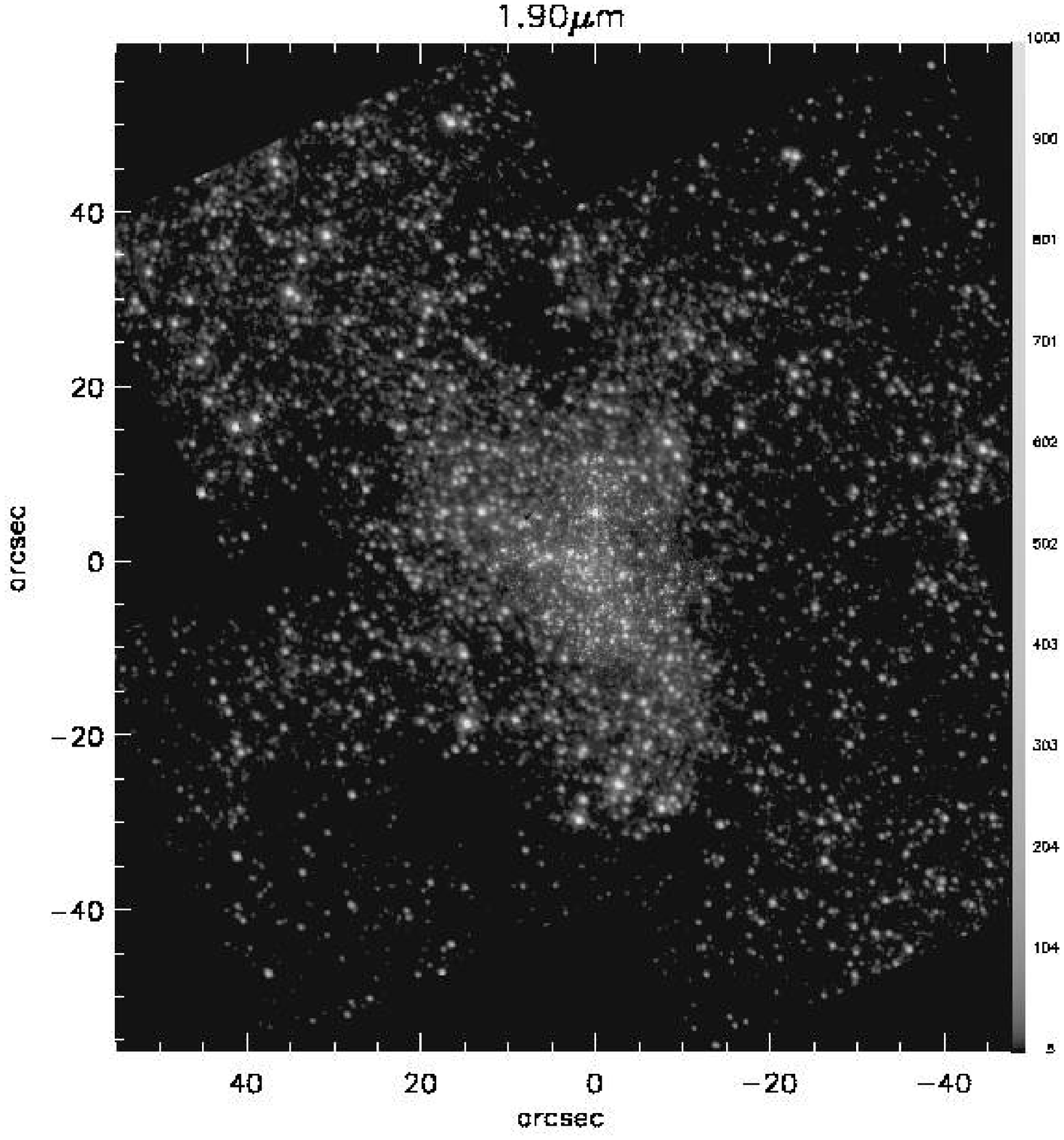}
\newpage
\plotone{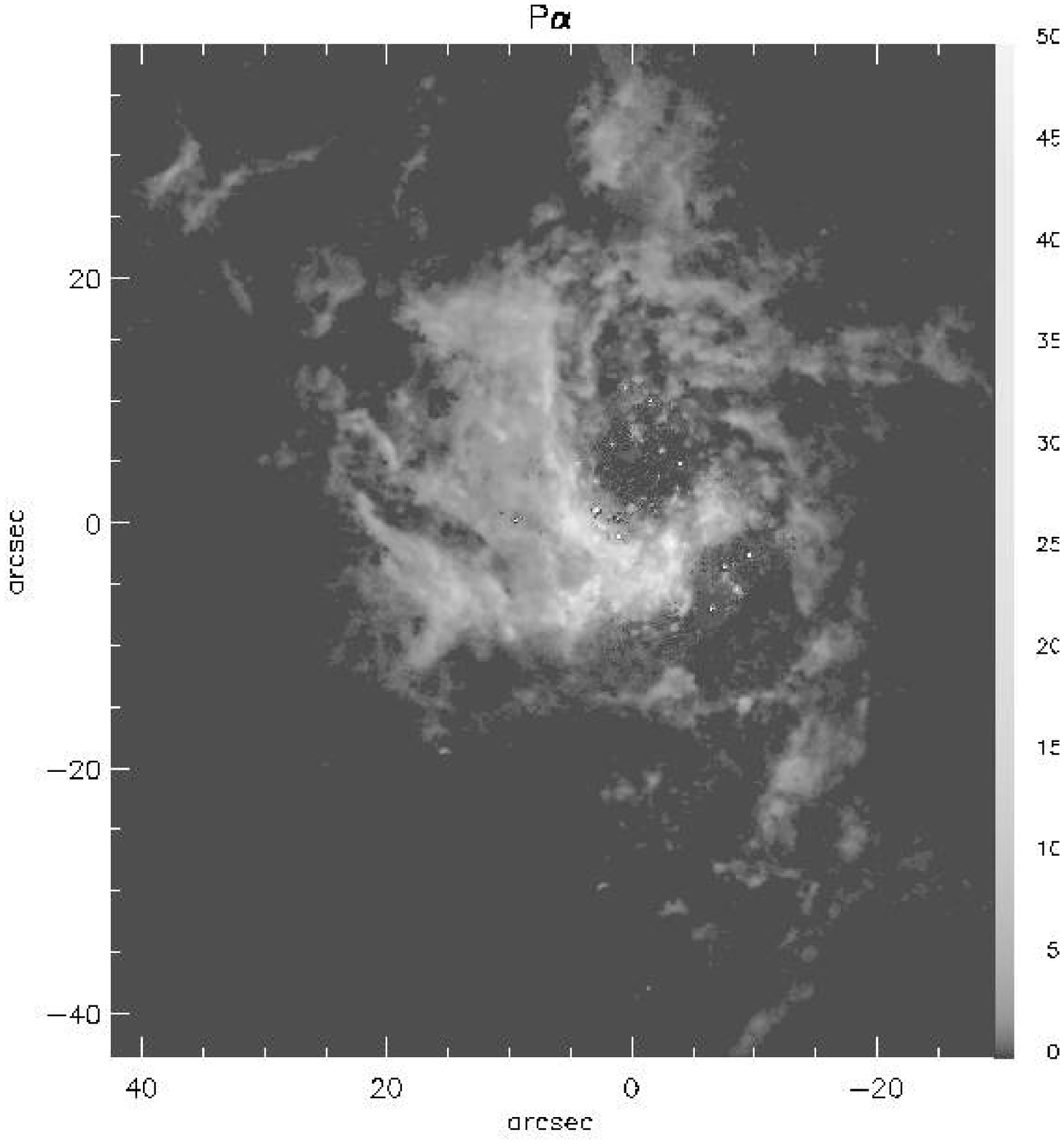}
\newpage
\plotone{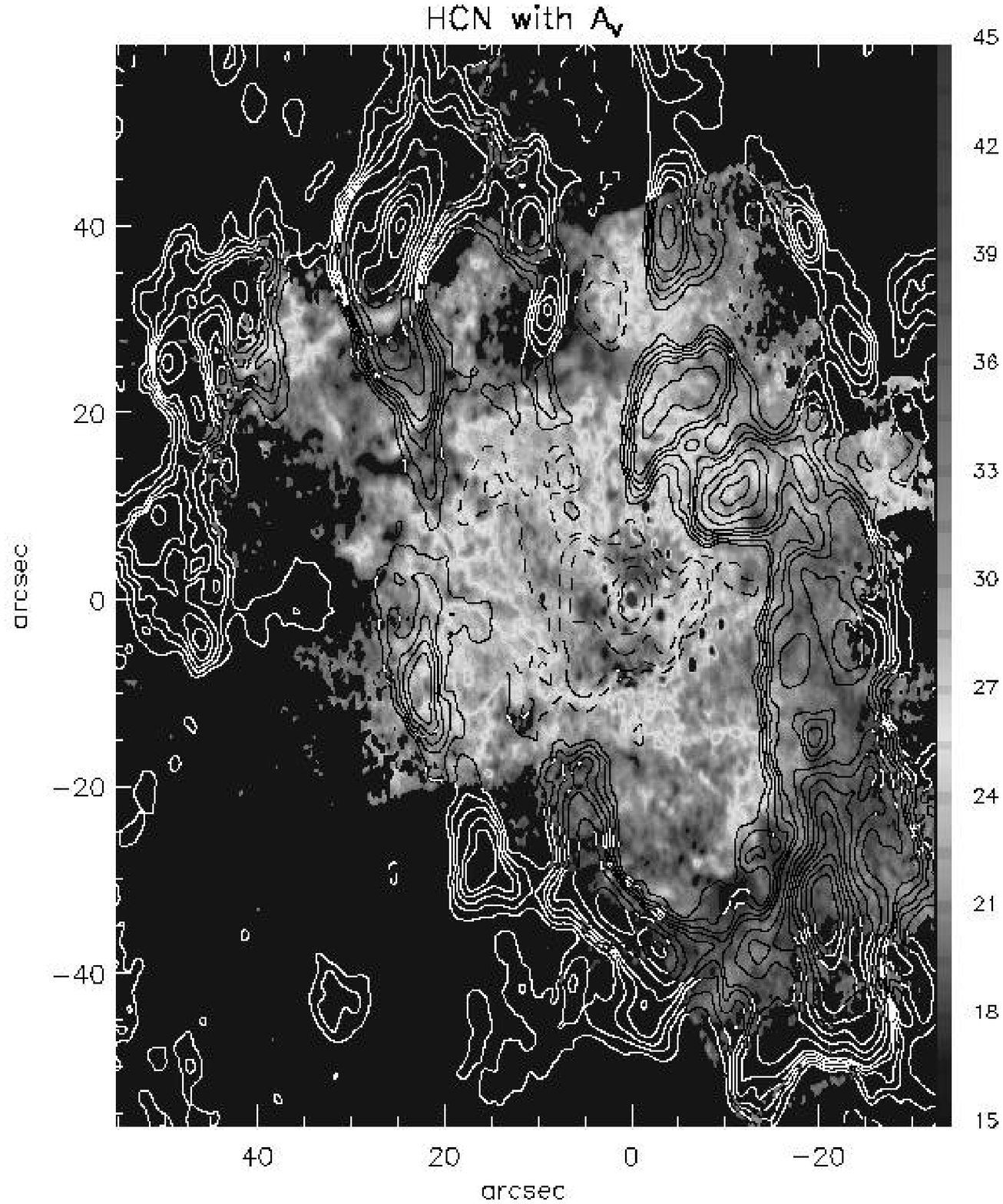}
\newpage
\plotone{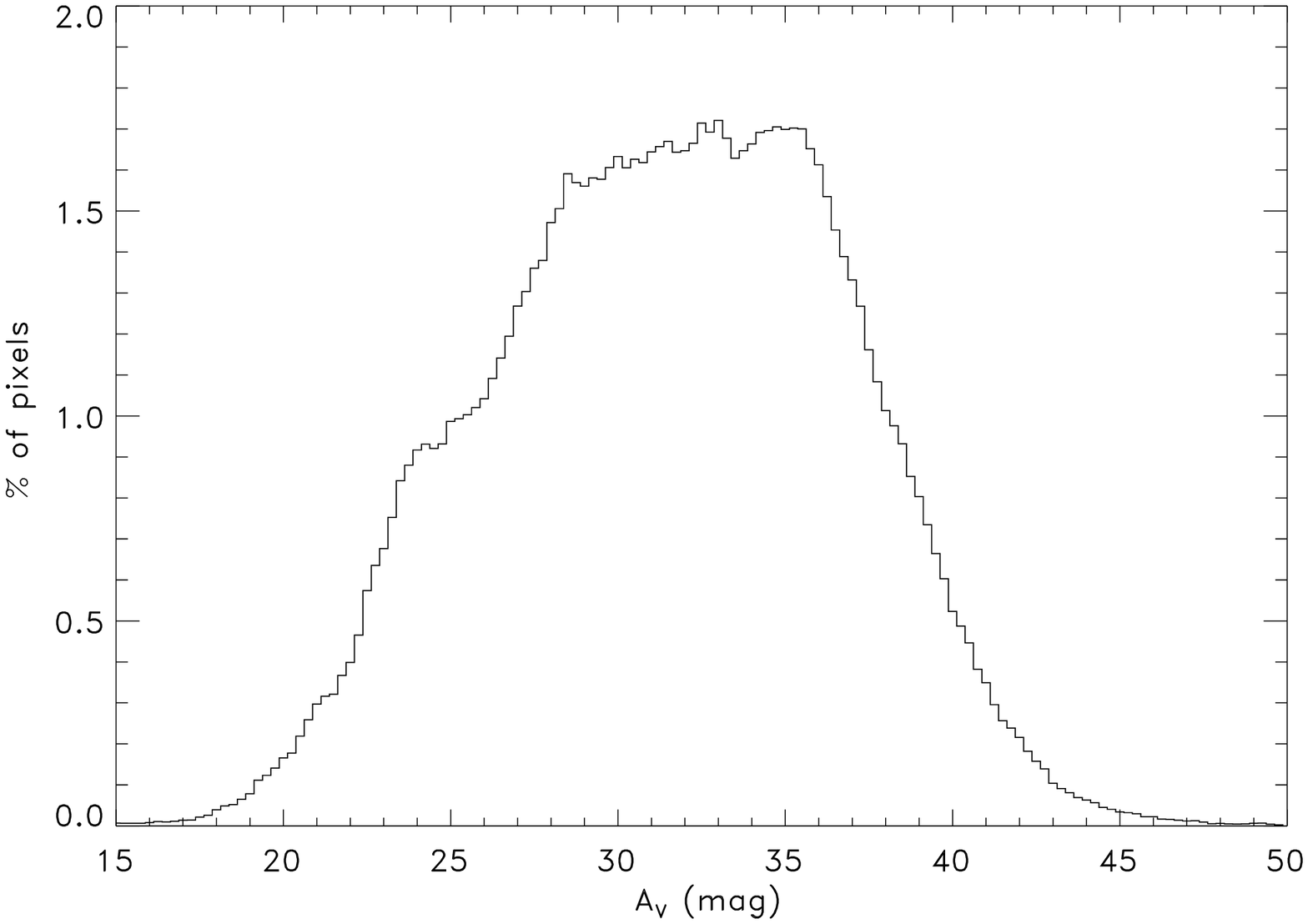}
\newpage
\plotone{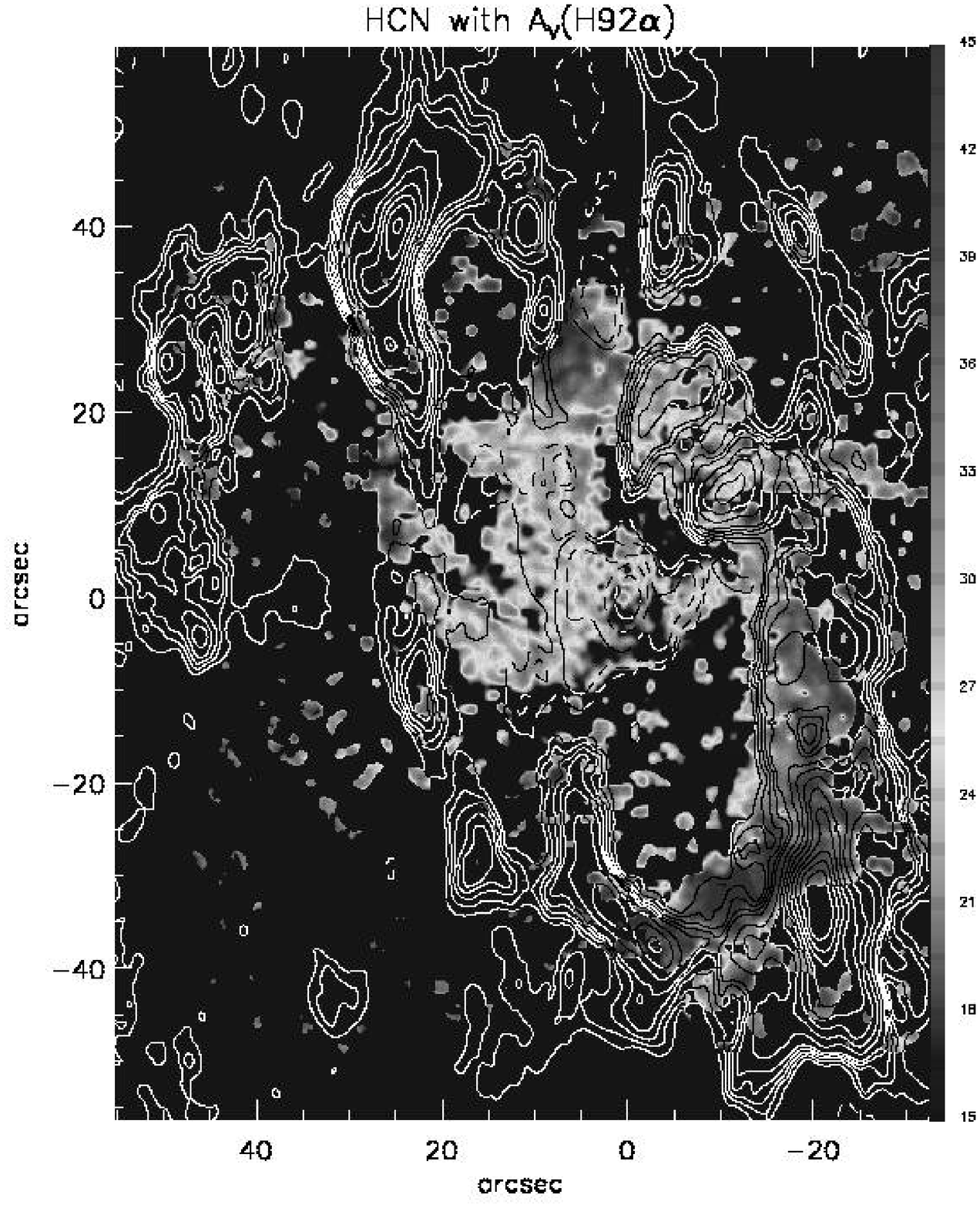}
\newpage
\plotone{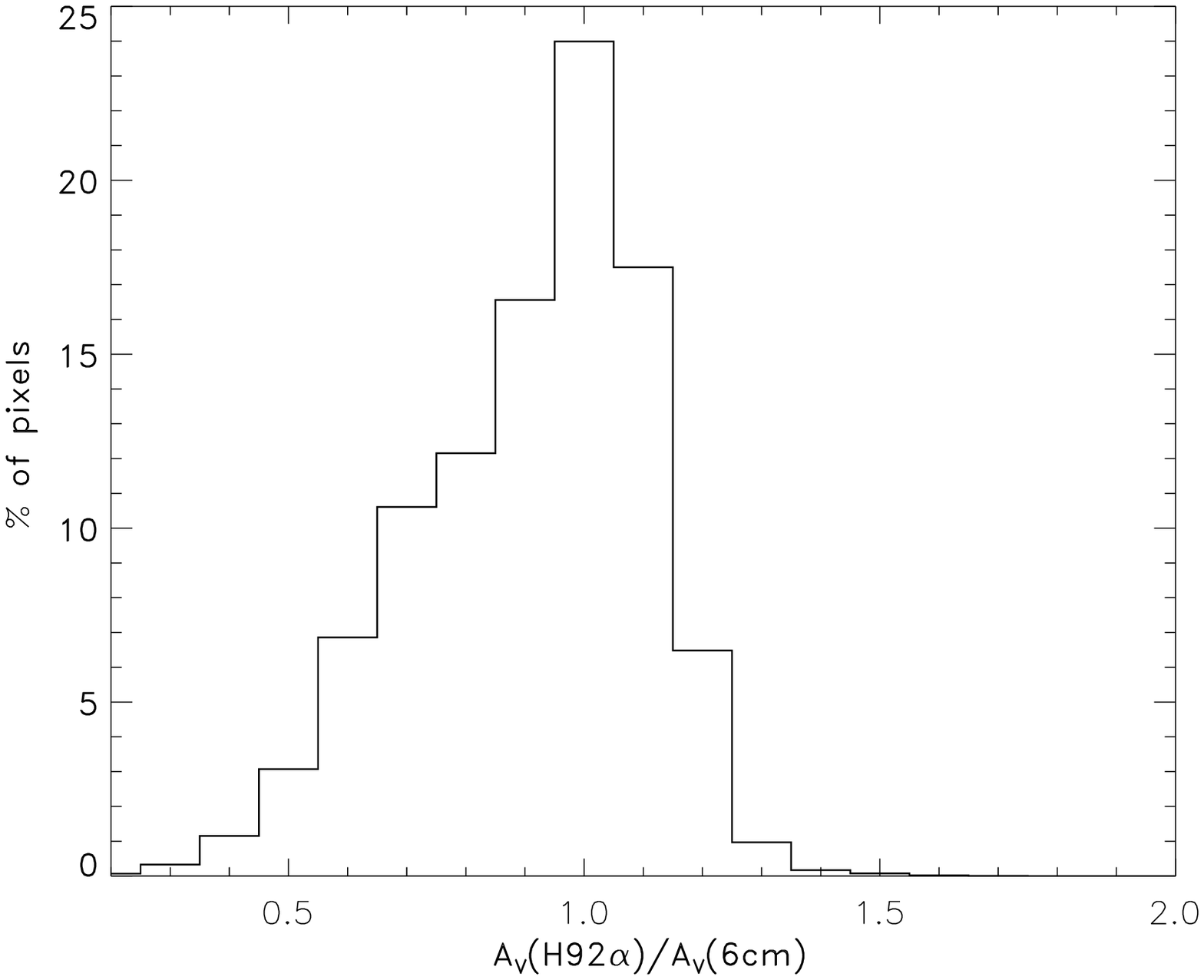}
\newpage
\plotone{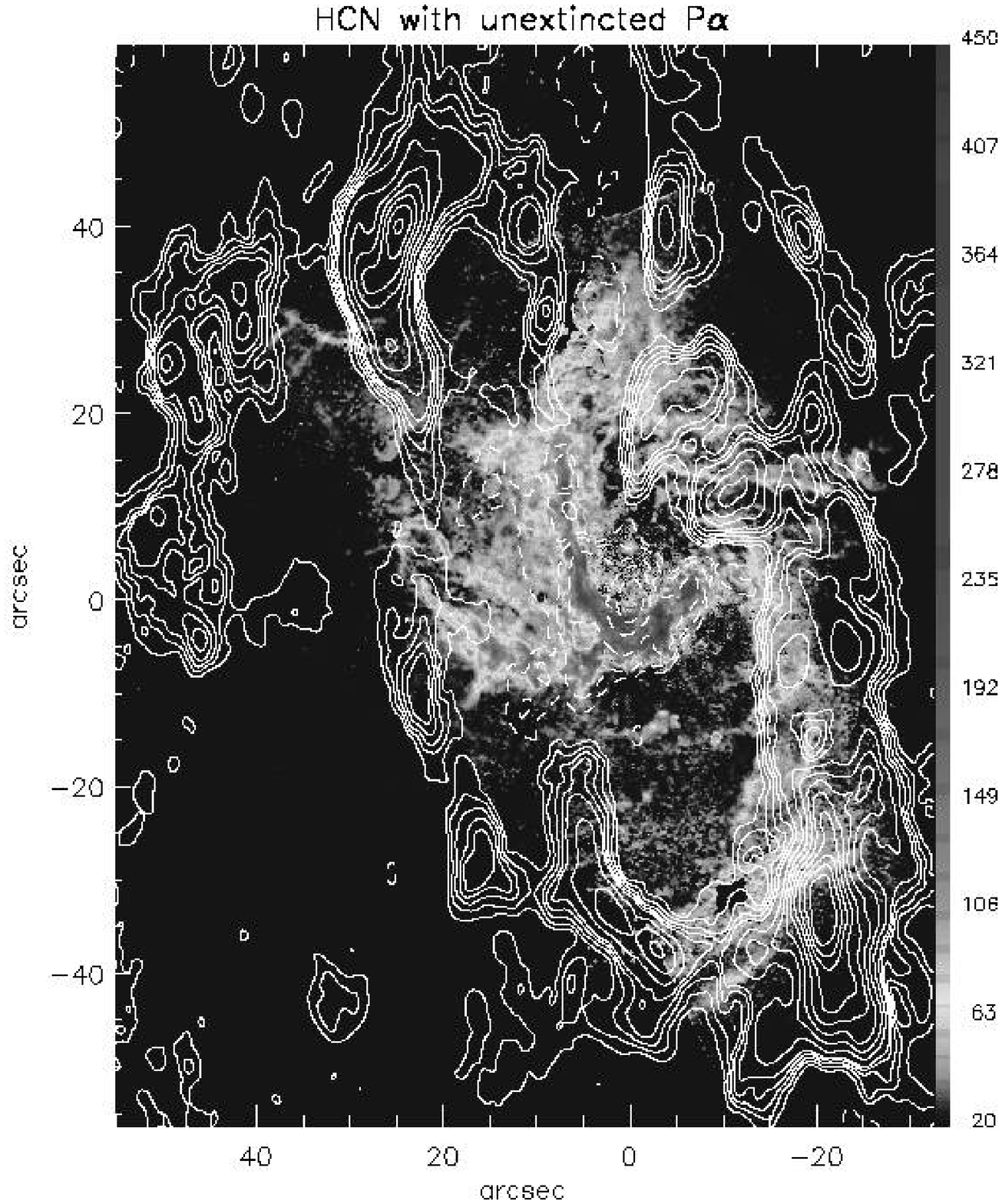}
\newpage
\plotone{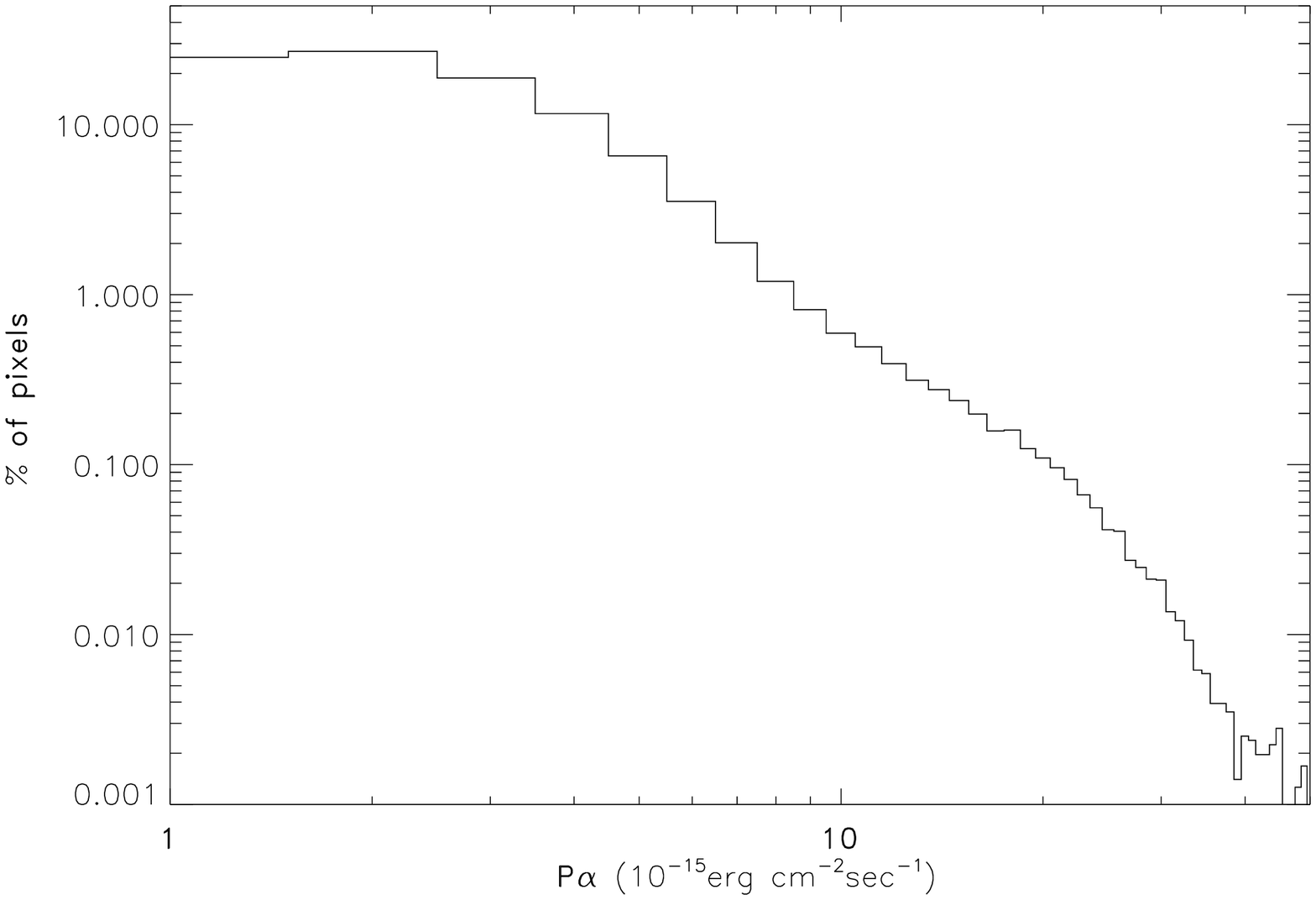}
\newpage
\plotone{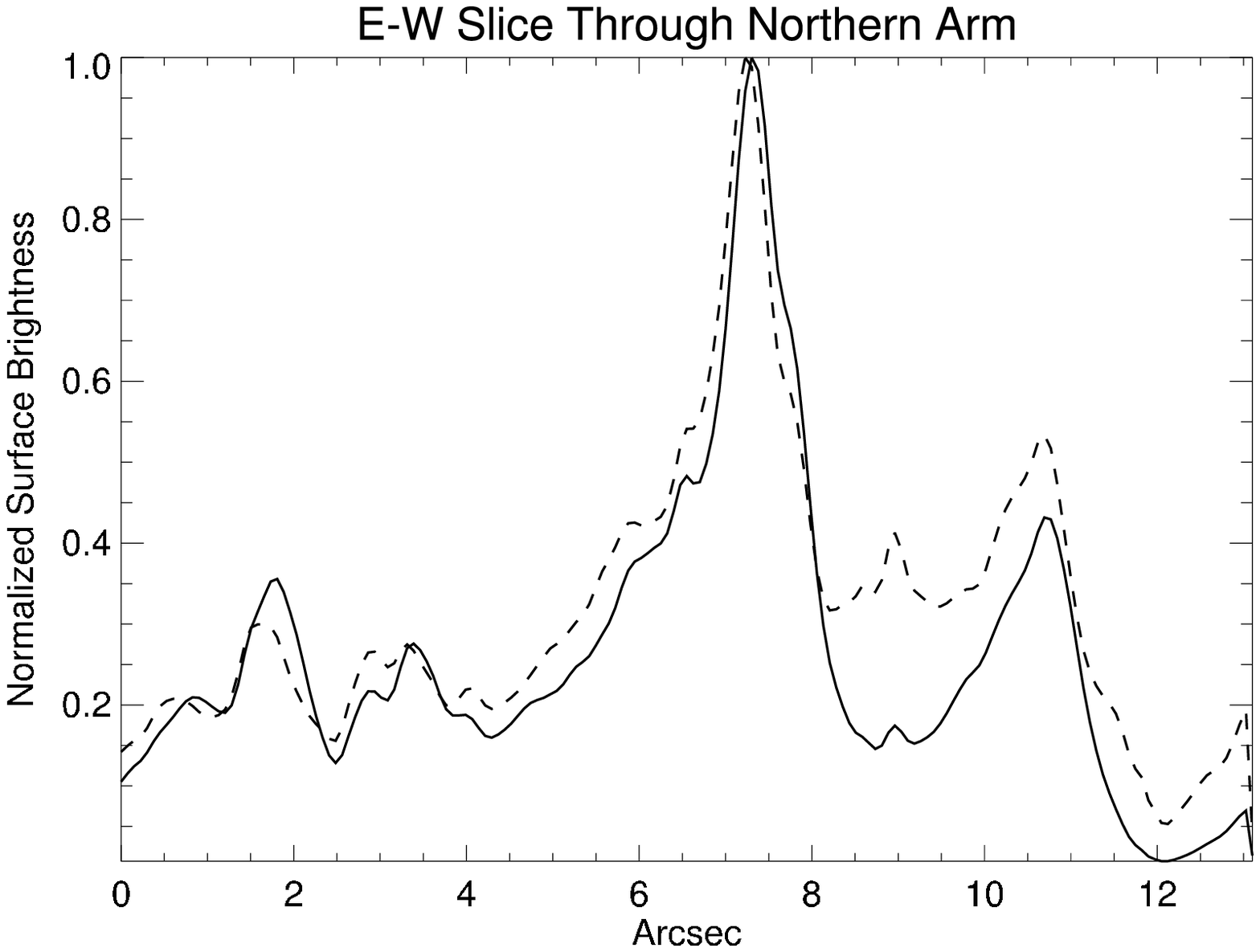}
\newpage
\plotone{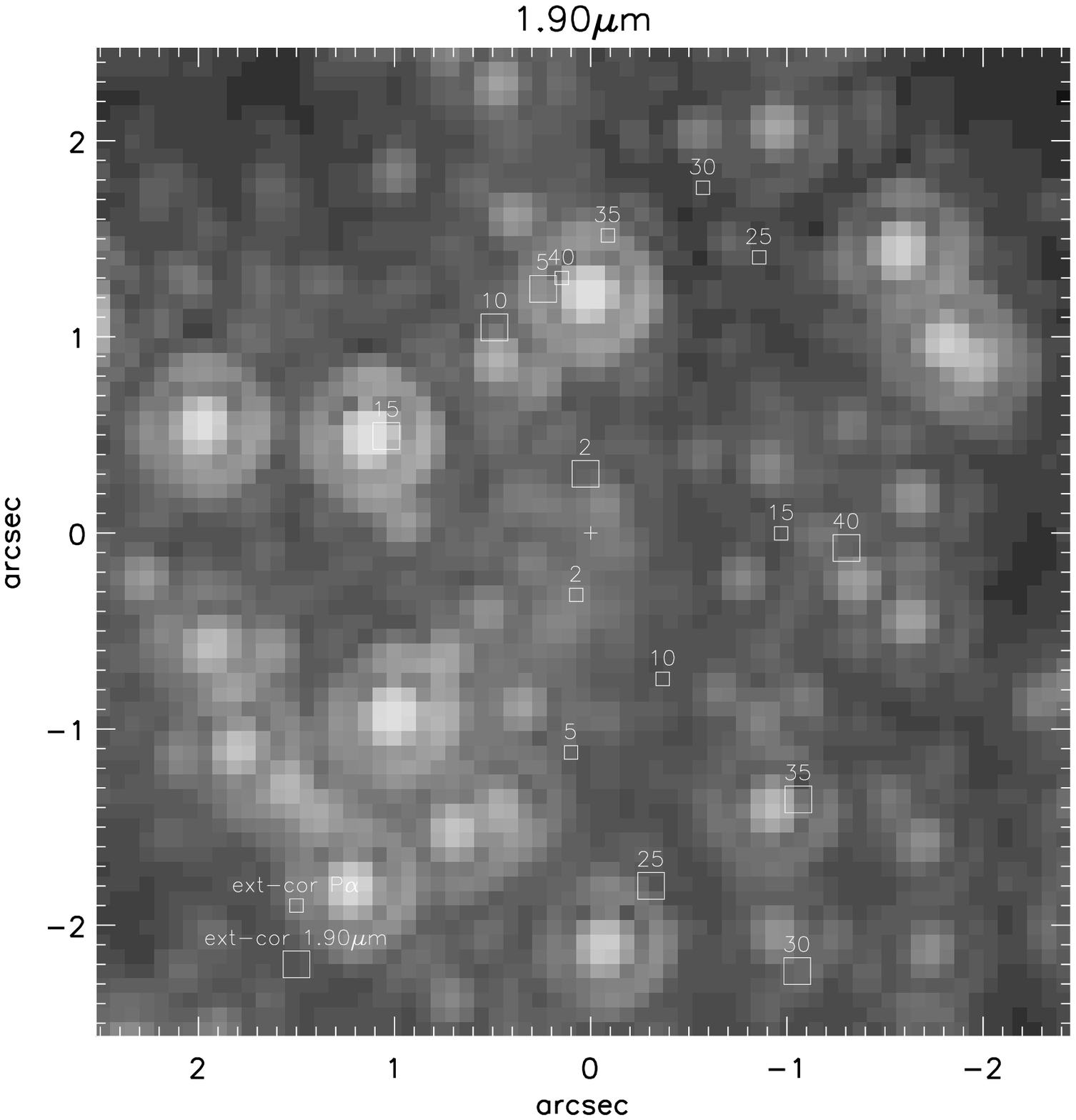}
\newpage
\plotone{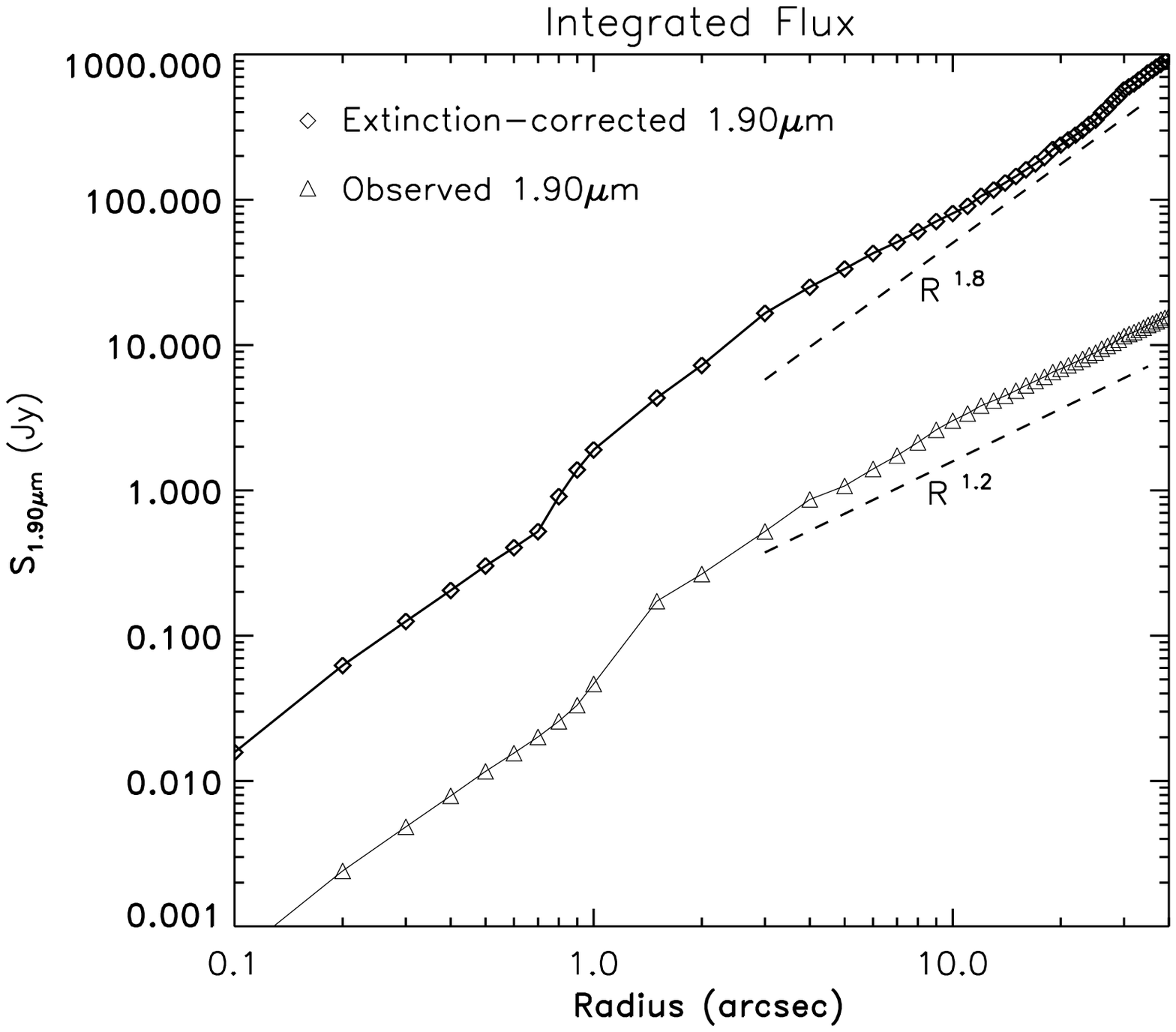}
\newpage
\plotone{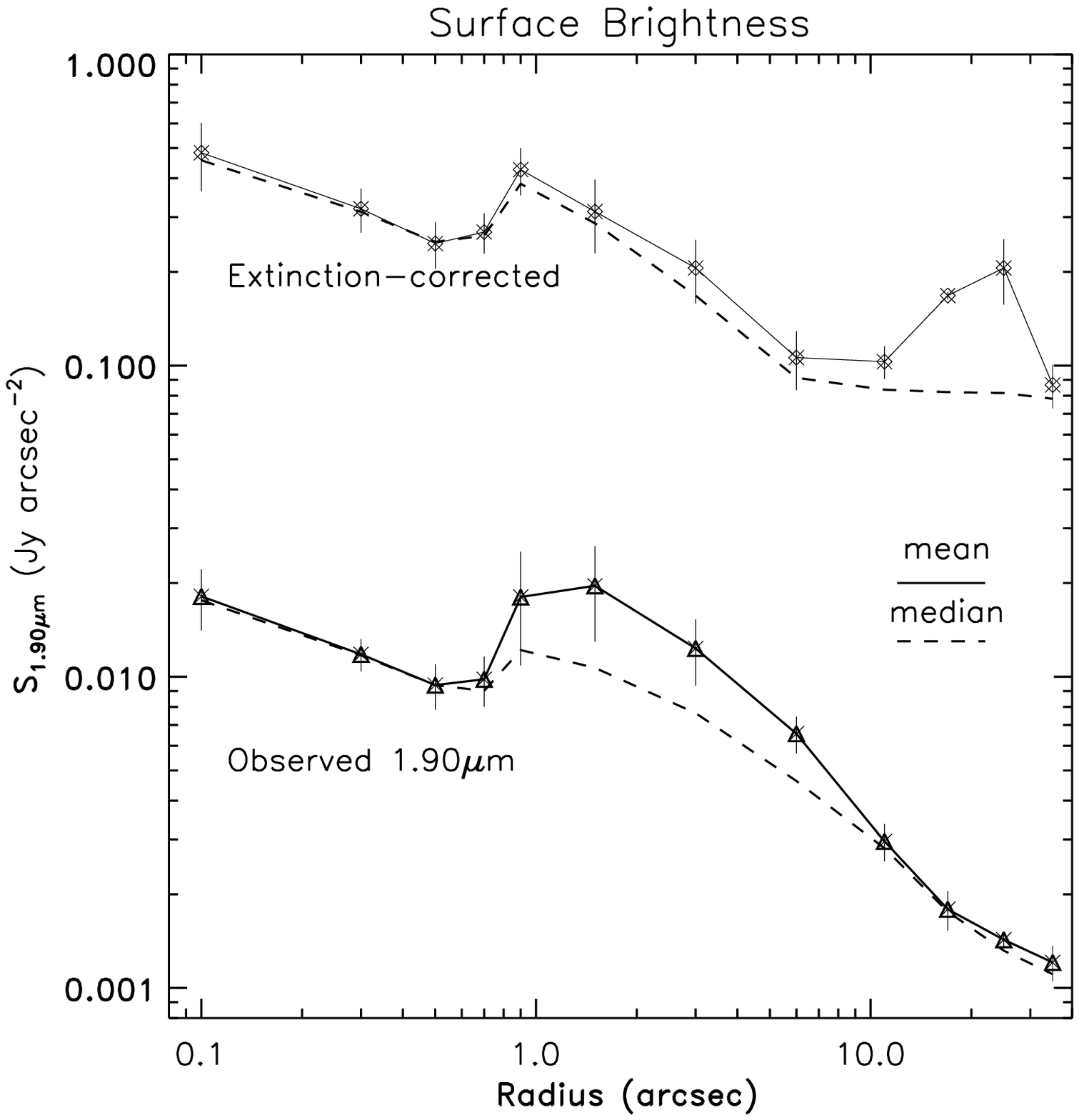}
\newpage
\plotone{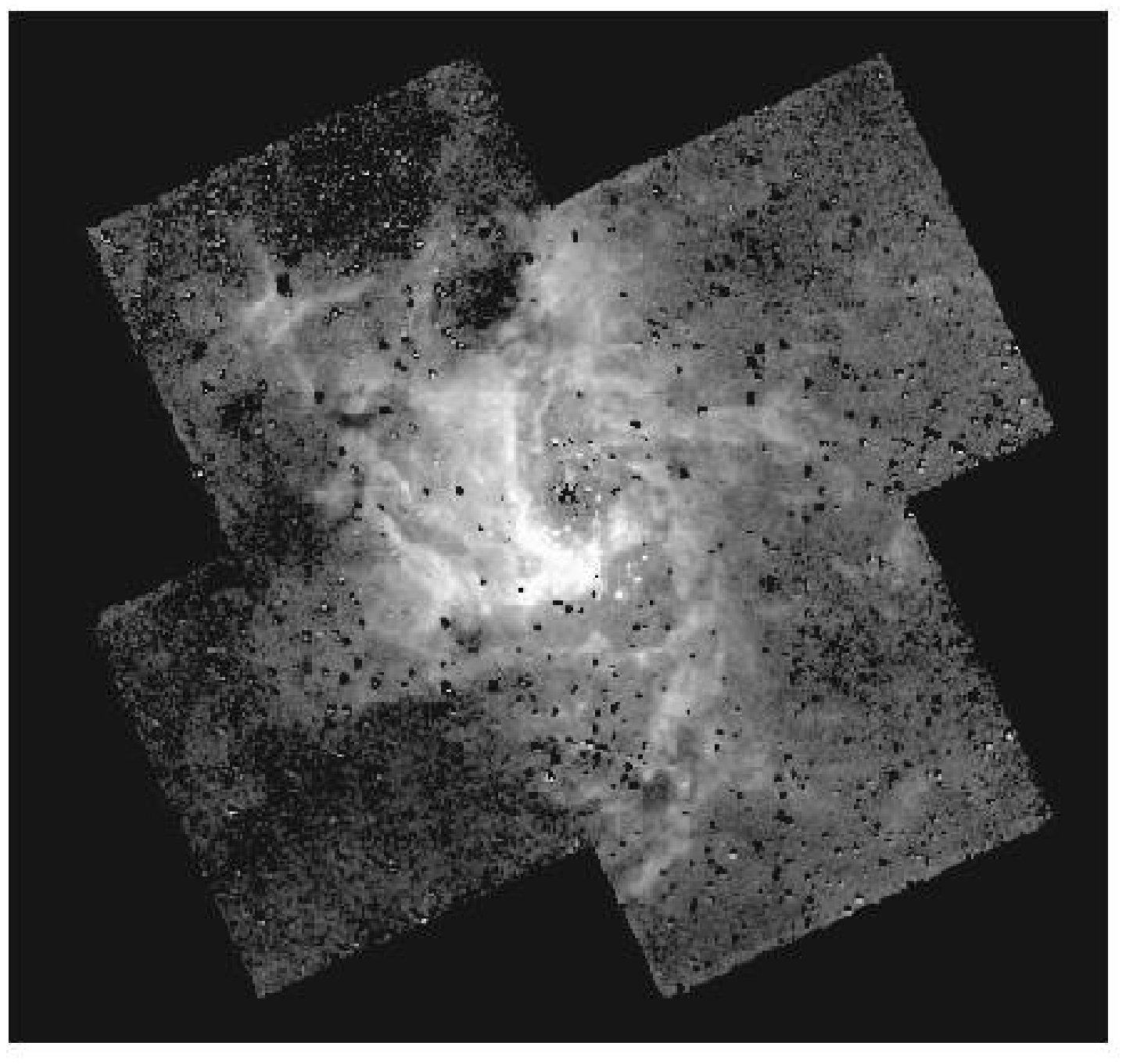}
\newpage
\plotone{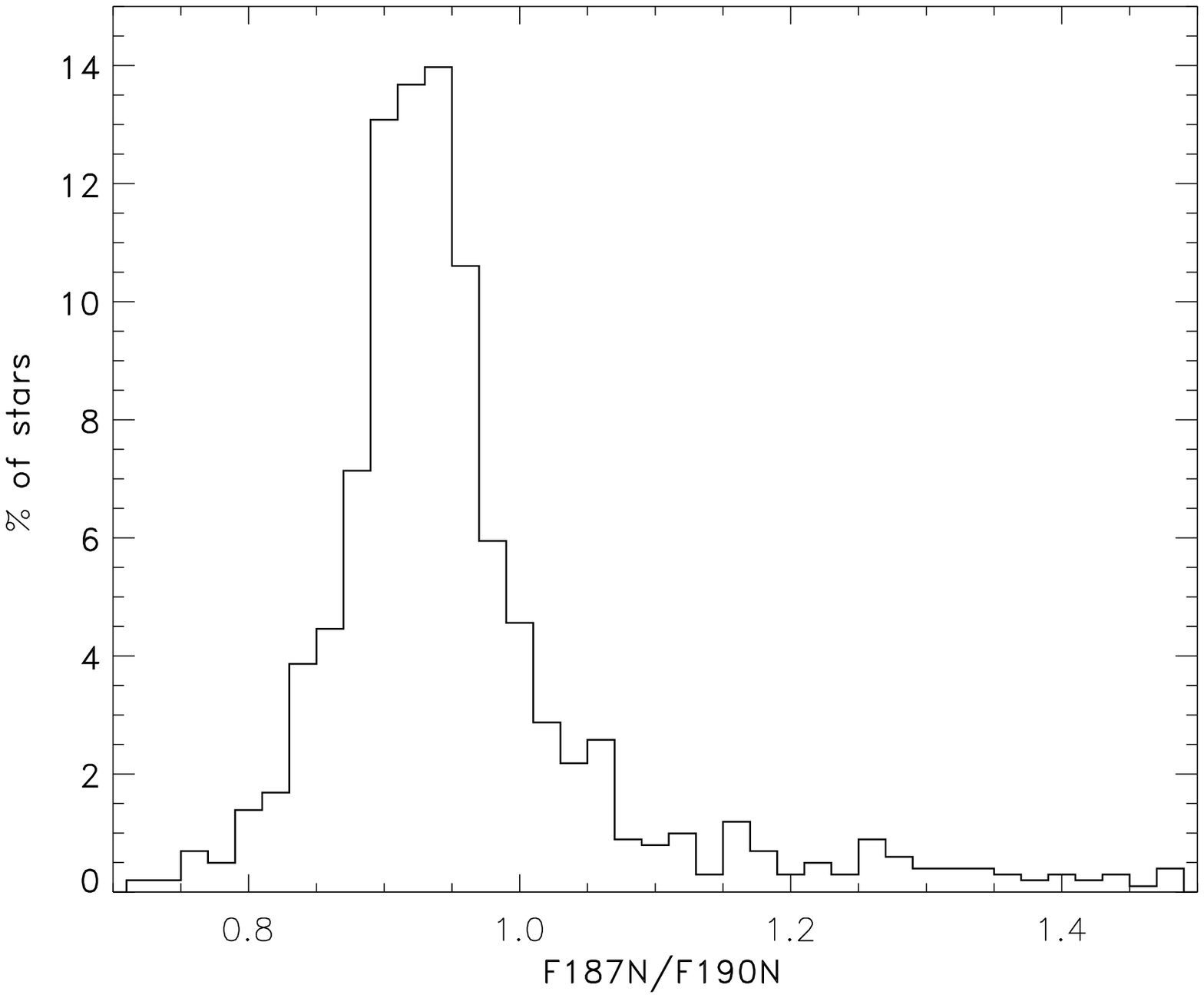}

\end{document}


\begin{deluxetable}{r r r r c c r l}
\footnotesize
\tablenum{2}
\tablewidth{0pt}
\tablecaption{P$\alpha$ Emission Line Stars in the Central Parsec\tablenotemark{a}}
\tablehead{ 
\colhead{~Source}&
\colhead{~~~~~$\Delta \alpha$ (\arcsec )\tablenotemark{b}}&
\colhead{~~~~~$\Delta \delta$ (\arcsec )\tablenotemark{b}}&
\colhead{~~~~~$\Delta r$ (\arcsec )\tablenotemark{b}}&
\colhead{~F187N/F190N \tablenotemark{c}}&
\colhead{~m$_{F190N}$ \tablenotemark{d}}& 
\colhead{~F$_{P\alpha}$ \tablenotemark{e}}& 
\colhead{~Identification \tablenotemark{f}} 
}
\label{datalog}
\startdata
  1 &   1.13	&   0.50 &   1.24   &  1.36  &  10.9    &   215   & IRS 16C\\
  2 &   0.00	&   1.24 &   1.24   &  1.20  &  11.0    &   107   & IRS 16NW\\
  3 &   1.00	&  -0.94 &   1.38   &  1.41  &  11.0    &   216   & IRS 16SW\\
  4 &  -0.97	&   2.09 &   2.31   &  1.10  &  13.3    &     7   & \\
  5 &   2.85	&   1.10 &   3.06   &  1.23  &   9.9    &   329   & IRS 16NE\\
  6 &   2.91	&  -1.16 &   3.13   &  2.59  &  13.1    &   126   & IRS 16SE2\\
  7 &   0.61	&  -3.09 &   3.15   &  1.21  &  11.0    &   106   & IRS 33SE\\
  8 &  -3.20 	&  -1.43  &   3.50  &  2.15  &  12.7    &   122   & IRS 13E4\\
  9 &  -3.15 	&  -1.72  &   3.59  &  2.07  &  11.9    &   236   & IRS 13E2\\
 10 &  -4.05	&   1.62 &   4.36   &  1.36  &  13.7    &      15   & IRS 34\\
 11 &   2.92	&   3.51 &   4.56   &  1.62  &  14.4    &      14   & HeIN3\\
 12 &  -4.37	&  -1.64 &   4.67   &  1.53  &  13.6    &      25   &\\
 13 &  -3.71	&   2.87 &   4.70   &  1.36  &  14.3    &	9   &\\
 14 &   0.91	&   5.81 &   5.88   &  1.18  &  12.7    &      19   &\\
 15 &   2.81	&  -5.59 &   6.26   &  2.07  &  13.0    &      89   & IRS 9W\\
 16 &  -3.95	&   4.95 &   6.33   &  1.63  &  13.1    &      47   & IRS 7W\\
 17 &  -2.46	&   6.01 &   6.49   &  1.80  &  14.3    &      19   &\\
 18 &   4.41	&   5.02 &   6.69   &  1.84  &  13.8    &      32   & IRS 7E2\\
 19 &  -7.63	&  -3.57 &   8.42   &  1.88  &  12.5    &   113   & AFNW\\
 20 &   4.22	&  -7.97 &   9.02   &  1.77  &  14.2    &      21  & \\
 21 &   9.47	&   0.37 &   9.48   &  1.92  &  13.6    &      43   & ID180\\
 22 &  -6.56	&  -6.92 &   9.54   &  3.09  &  11.0    &   479   & AF\\
 23 &  -9.64	&  -2.57 &   9.98   &  2.38  &  13.3    &      90   &\\
 24 &  -8.64	&  -5.36 &  10.17   &  1.75  &  13.8    &      32   & \\
 25 &  -1.53	&  10.05 &  10.17   &  1.78  &  12.7    &      81   & IRS 15SW\\
 26 &   1.24	&  11.81 &  11.88   &  1.38  &  12.0     &   80   & IRS 15NNE\\

\enddata
\tablenotetext{a}{Stars for which the flux in F187N is $\geq$ 10\%
more than in F190N (where the F190N image was multiplied by a factor of 0.91 as discussed in the text)}
\tablenotetext{b}{Offsets and radius are relative to 
\astar ($\alpha_{2000}$ = 17$^h$45$^m$40$^s$.058, 
$\delta_{2000}$ = -29\deg00\arcmin27.90\arcsec).}
\tablenotetext{c}{F187N/F190N flux ratio}
\tablenotetext{d}{F190N flux density (magnitude), not corrected for extinction; zero magnitude = 808 Jy. 
All fluxes extracted from NIC2 images using ``starfinder'' with the exceptions of IRS 13E4 and
IRS 13E2 (which required PSF subtraction) and IRS 15NNE (which appears only in the NIC3 image,
and for which aperture photometry was performed).}
\tablenotetext{e}{Units : 10$^{-15}$ ergs cm$^{-2}$ sec$^{-1}$. 
The fluxes have not been corrected for extinction.}
\tablenotetext{f}{Identifications with stars listed by Genzel \etal (2000),
Ott, Eckart \& Genzel (1999), Paumard \etal (2001), and Maillard \etal (2003). 
Our positions were checked against those given by Menten \etal (1997) 
and we find agreement to within 0.15\arcsec\ in all cases.} 
\end{deluxetable}